\documentclass[usenatbib]{mn2e}
\usepackage{graphicx}
\title[Vortices in self-gravitating gaseous discs]
{Vortices in self-gravitating gaseous discs}
\author[G.~R.~Mamatsashvili and W.~K.~M.~Rice] {G.~R.~Mamatsashvili$^{1,2}$\thanks{E-mail:
grm@roe.ac.uk} and W.~K.~M.~Rice
$^{1}$\\
$^{1}$ SUPA, Institute for Astronomy, University of Edinburgh,
Blackford Hill, Edinburgh EH9 3HJ, Scotland \\
$^{2}$ Georgian National Astrophysical Observatory,
Il. Chavchavadze State University, 2a Kazbegi Ave., Tbilisi 0160, Georgia}
\begin{document}

\date{Accepted 2009 January 09. Received 2009 January 08; in original form 2008
November 06}

\pagerange{\pageref{firstpage}--\pageref{lastpage}} \pubyear{2009}

\maketitle

\label{firstpage}

\begin{abstract}
Vortices have recently received much attention in the research of
planet formation, as they are believed to play a role in the
formation of km-sized planetesimals by collecting dust particles in
their centres. However, vortex dynamics is commonly studied in
non-self-gravitating discs. The main goal here is to examine the
effects of disc self-gravity on vortex dynamics. For this purpose,
we employ the 2D shearing sheet approximation and numerically solve
the basic hydrodynamic equations together with Poisson's equation to
take care of disc self-gravity. A simple cooling law with a constant
cooling time is adopted, such that the disc settles down into a
quasi-steady gravitoturbulent state. In this state, vortices appear
as transient structures undergoing recurring phases of formation,
growth to sizes comparable to a local Jeans scale and eventual
shearing and destruction due to the combined effects of self-gravity
(gravitational instability) and background Keplerian shear. Each
phase typically lasts about 2 orbital periods or less. As a result, in
self-gravitating discs the overall dynamical picture of vortex
evolution is irregular consisting of many transient vortices at
different evolutionary stages and, therefore, with various sizes up
to the local Jeans scale. By contrast, in the non-self-gravitating
case, long-lived vortex structures persist for hundreds of orbits
via merging of smaller vortices into larger ones until eventually
their size reaches the disc scale height. Vortices generate density
waves during evolution, which turn into shocks. This phenomenon of
wave generation by vortices is an inevitable consequence of the
differential character (shear) of disc rotation. Therefore, the
dynamics of density waves and vortices are coupled implying that, in
general, one should consider both vortex and spiral density wave
modes in order to get a proper understanding of self-gravitating
disc dynamics.

Our results suggest that given such an irregular and rapidly varying
character of vortex evolution in self-gravitating discs, it may be
difficult for such vortices to effectively trap dust particles in
their centres, a proposed mechanism for planetesimal formation.
Further study of the behaviour of dust particles embedded in a
self-gravitating gaseous disc is, however, required to strengthen
this conclusion.
\end{abstract}

\begin{keywords}
accretion, accretion discs -- gravitation -- hydrodynamics --
instabilities -- (stars:)planetary systems: protoplanetary discs -- turbulence
\end{keywords}

\section{Introduction}

Research in protoplanetary disc dynamics is mainly focused on
investigating the dynamical activities of two basic types/modes of
perturbations -- spiral density waves and vortices. They provide a
means of outward angular momentum transport necessary for
the secular evolution of a neutral disc, where
the magnetorotational instability cannot
operate. However, each mode is analysed for different
settings: spiral density waves are the central perturbation types
and, therefore, the main agents responsible for angular momentum
transport in self-gravitating discs, while vortices are commonly
studied in non-self-gravitating discs. The formation and dynamics of
vortices in protoplanetary discs has only recently attracted
attention because of the role they may play in planet formation. The
latest developments have revealed that in non-self-gravitating discs vortices can be (linearly)
coupled with spiral density waves.
This noteworthy finding, in turn, calls for revisiting
self-gravitating disc dynamics with a particular emphasis on the
possible role of vortical perturbations in the overall dynamical
picture together with spiral density waves. Below we first outline
the recent results on the vortex dynamics in astrophysical discs.

It has been demonstrated in both 2D global
\citep{Br99,GL99a,GL99b,GL00,DSC00,Lietal01,Da02,KB03,Bo07} and
local shearing sheet \citep{UR04,BM05,JG05,SSG06,L07} simulations of
non-self-gravitating discs that only anticyclonic vortices (rotating
in the same sense as the background shear) can survive in discs,
though due to compressibility and/or viscosity they slowly decay on
the timescale of several hundred orbital periods. Nevertheless,
anticyclonic vortices can be thought of as long-lived structures. On
the other hand, cyclonic vortices get strongly sheared by
differential rotation of the disc and eventually disappear.
Simulations are often initiated either with random potential
vorticity (PV) perturbations, which contain positive (cyclonic) and
negative (anticyclonic) values in equal portions, or with an imposed
single vortex. The former shows that initial small scale regions
where PV is negative (anticyclonic regions) first get sheared into
strips and then start to wrap up into small scale vortices due to
the specific instability discussed below. The latter gradually grow
in size via merging into larger vortices until eventually their size
becomes of order the disc scale height \citep{JG05}; growth beyond
this scale is restricted by compressibility effects. The initial
small scale cyclonic regions instead of wrapping up into distinct
vortices get strongly sheared into strips and remain so during the
entire course of evolution. This limitation on the vortex size was
also confirmed in detailed simulations of a single vortex in a
Keplerian disc \citep{Bo07}. In these simulations, a vortex with an
initial length exceeding the disc scale height undergoes nonlinear
adjustment, during which it decreases in size radiating excess
energy in the form of spiral density waves and shocks, and finally
settles down to scales equal to a few disc scale height. These final
scales are independent of the initial vortex size and are determined
by disc properties (sound speed and disc scale height).

The emission of spiral density waves by vortices during the
adjustment process is due to the background Keplerian shear that
ensures (linear) coupling between these two modes when the
horizontal scale of vortices is equal to or larger than the disc
scale height \citep{Lietal01,Da02,JG05,Bo05,Bo07,MC07}. On the other
hand, this condition is nothing more than the validity of 2D
treatment. So, in the 2D case compressibility is a key feature that
must be taken account of. In this respect, all the results of 2D
incompressible simulations of vortices should be regarded as
approximate and less realistic. Emitted spiral density waves steepen
into shocks afterwards and, hence, vortices appear to generate
shocks that may play an important role in the outward transport of
angular momentum, particularly in neutral discs where the
magnetorotational instability cannot operate \citep[see
e.g.][]{Lietal01,Da02,JG05}.

The basic underlying mechanism/instability responsible for the
development of vortices as well as the necessary criterion for that
were identified by \citet{L07} for incompressible shear flows. He
interprets it as nonlinear Kelvin-Helmholtz instability of a vortex
strip with vorticity of the same sign as the background vorticity
(for incompressible flows vorticity and PV are equivalent); for a
vortex strip with the opposite sign of vorticity no such instability
is observed. This readily explains why cyclonic vortices cannot
survive. This mechanism of vortex formation should be extended to
compressible shear flows as well. Note that in this scenario one
needs PV perturbations (e.g., vortex strips or random type)
initially present in a disc. Other mechanisms of vortex, or PV,
generation that do not necessarily require this include
inhomogeneities of entropy (temperature) distribution of a disc and
are known as baroclinic instability \citep{KB03,K04,Petal07} and
Rossby wave instability \citep{Lo99,Lietal01}.

Another important property of anticyclonic vortices -- their dust
trapping capability -- was also investigated
\citep[e.g.,][]{BS95,Ch00,GL00,dFMB01,JAB04,IB06,KB06}. It has been
shown that a smooth, sufficiently long-lived vortex is indeed able
to effectively trap dust particles in its core, possibly
accelerating planetesimal formation.

All the above-mentioned studies on vortex dynamics miss out an
important aspect of protoplanetary discs -- self-gravity (see
reviews by \citealt{AL93}, \citealt{Duetal07} and \citealt{Lod07} on
the role of self-gravity in protoplanetary disc dynamics). Typically
in discs effective cooling times are too long to cause fragmentation
under the action of their own self-gravity \citep{Boletal06}. As a
consequence, balance is established between heating due to
gravitational instability and cooling (self-regulation mechanism,
\citealt{BL01}). Discs are expected to stay in this self-regulated
quasi-steady gravitoturbulent state for a long time. In this case
Toomre parameter hovers on the margin of gravitational instability.
The properties of this state were numerically investigated in a
large number of papers \citep[e.g.,][and references
therein]{LB94,LKA97,NBR00,G01,JG03,Rietal03,LR04,LR05,Me05,Boletal06,SW08}.
A general dynamical picture is that spiral structure develops in a
disc and transports angular momentum outwards through gravitational
and hydrodynamic stresses, thereby allowing matter to accrete onto
the central star. The angular momentum transport in this case is
attributed to spiral density waves, which are thought of as the only
perturbation type present in the disc. In other words, almost all
studies on self-gravitating disc dynamics concentrate on the
dynamical activity of spiral density waves and leave another class
of perturbations -- vortices -- out of consideration. As discussed
above, the latter plays an important role in non-self-gravitating
discs and it seems natural to look for them and analyse their
nonlinear development in self-gravitating discs too. Indeed, in the
linear theory it has already been shown that the coupling between
spiral density wave and vortex modes is even more efficient in the
presence of self-gravity \citep{MC07}. In perspective, this study
will allow us to see if the same mechanism of planetesimal formation
-- dust particle trapping by vortices -- can also be at work in
self-gravitating discs.

In order to study the dynamics of vortical perturbations in
self-gravitating discs, one must examine the behaviour of the PV
field -- the basic quantity characterizing vortex development. To
date no systematic investigation of PV behaviour, similar to that
done for non-self-gravitating discs, has been carried out for
self-gravitating ones. However, we should mention two relevant works
by \citet{AW95} and \citet{WMN02}.

\citet{AW95} investigated the dynamics of a single vortex in the
quasi-geostrophic and local shearing sheet approximations in a
self-gravitating Keplerian disc. The quasi-geostrophic approximation
implies that the characteristic timescale of a problem is much
larger than the orbital period. The vortex considered in their paper
is in geostrophic balance and, therefore, remains steady for many
rotation periods. The gas motion inside the vortex is subsonic as
well. In this case the effect of self-gravity is only to make the
effective length scale (Rossby radius) of the vortex larger than
that in the non-self-gravitating case. The quasi-geostrophic
approximation does not permit consideration of the most important
aspect of dynamics -- compressibility effects (spiral density waves
and shocks), which are intertwined with vortices and have typical
timescales of the order of orbital (shear) time. In other words,
these relatively fast motions associated with compressibility are
filtered out in this approximation.

\citet{WMN02} investigated the properties of the gravitoturbulent
state in the interstellar medium of galaxies. The vorticity (but not
PV) field in this state is indeed calculated in their paper, which
has a rather complicated structure. However, the authors do not
discuss in detail the properties of the vorticity field and its
relation to the density and pressure fields. They only point out
that negative and positive vorticity regions are associated with
dense filaments seen in the density field. In that paper the main
emphasis is placed on analysing the spectral properties of
gravitoturbulence. Their study clearly demonstrates that vortical
perturbations are as important as spiral density waves in the
formation of spectra of the resulting gravitoturbulent state.

In the present paper, following the approach adopted by \citet{G01}
and \citet{JG05}, we study the specific properties of vortex, or PV,
evolution in self-gravitating discs by means of numerical
simulations. We work in the 2D shearing sheet approximation
\citep{GLB65,GT78} without invoking the quasi-geostrophic
approximation, thereby allowing for compressibility effects. In this
respect our analysis is more general than that of \citet{AW95}. In
addition, we do not impose a single vortex in quasi-equilibrium
balance in the beginning. Instead, we start with random
perturbations of velocity components and, hence of PV, and trace the
development of structures out of this chaotic field. These chaotic
initial perturbations are more realistic than a single vortex.
\emph{The main focus here is on the dynamical picture of PV
evolution in the state of quasi-steady gravitoturbulence and how
this picture differs from that occurring in non-self-gravitating
discs given the same chaotic type of initial conditions}. This, in
turn, allows us to draw important conclusions about the effects of
self-gravity on the formation and evolution of vortices. We will see
that in the presence of self-gravity, vortex evolution is not as
smooth and regular as it is in the non-self-gravitating case. The
present study is actually an extension to the nonlinear regime with
added cooling of \citet{MC07}, where we carried out linear analysis
of the transient growth and coupling of vortices and spiral density
waves in self-gravitating discs.

The paper is organised as follows. Physical approximations and the
mathematical formalism of the problem as well as the numerical
techniques used are introduced in Section 2, nonlinear evolution of the
fiducial model and the comparative study of correlation functions
for different models are described in Section 3 and summary and
discussions are given in Section 4.

\section{Physical Model and Equations}

In order to study the evolution of vortices in thin self-gravitating
gaseous discs, we employ the 2D local shearing sheet model. This
model represents a simple analogue to a differentially rotating disc
and has an advantage over global disc models in that it permits
higher resolution study of dynamical processes in discs. In the
shearing sheet model, disc dynamics is studied in the local
Cartesian coordinate frame corotating with the angular velocity of
disc rotation at some radius from the central star. In this
coordinate frame, the unperturbed differential rotation of the disc
manifests itself as a parallel azimuthal flow ${\bf u_0}$ with a
constant velocity shear in the radial direction. The unperturbed
background surface density $\Sigma_0$ and two-dimensional pressure
$P_0$ corresponding to this shear flow are assumed to be spatially
constant. A Coriolis force is also included to take into account the
effects of the coordinate frame rotation. As a result, in this local
approximation the continuity equation and equations of motion take
the form \citep{GT78,G01}:
\begin{equation}
\frac{\partial \Sigma}{\partial t}+\nabla\cdot(\Sigma{\bf u})
-q\Omega x \frac{\partial \Sigma}{\partial y} = 0,
\end{equation}
\begin{equation}
\frac{\partial u_x}{\partial t}+({\bf u}\cdot\nabla)u_x-q\Omega x
\frac{\partial u_x}{\partial y} =-\frac{1}{\Sigma}\frac{\partial
P}{\partial x}+2\Omega u_y-\frac{\partial \psi}{\partial x},
\end{equation}
\begin{equation}
\frac{\partial u_y}{\partial t}+\left({\bf u}\cdot\nabla \right)u_y
-q\Omega x \frac{\partial u_y}{\partial y}
 =-\frac{1}{\Sigma}\frac{\partial P}{\partial y}+(q-2)\Omega
 u_x-\frac{\partial \psi}{\partial y}.
\end{equation}
This set of equations is supplemented with
Poisson's equation for a razor-thin disc
\begin{equation}
\Delta \psi=4\pi G (\Sigma-\Sigma_0)\delta(z).
\end{equation}
Here ${\bf u}(u_x, u_y), P, \Sigma$ and $\psi$ are, respectively,
the perturbed velocity relative to the background parallel shear
flow ${\bf u_0}(0, -q\Omega x)$, the two-dimensional pressure, the
surface density and the gravitational potential of the gas sheet.
Since equations (2-3) are written for perturbed velocities, only the
gravitational potential due to the perturbed surface density
$\Sigma-\Sigma_0$ is used. In the dynamical equations, the gradients
of the gravitational potential are taken at $z=0$, i.e., where the
shearing sheet is located, because only this quantity depends on the
vertical coordinate $z$. $\Omega$ is the angular velocity of the
coordinate frame rotation as a whole. $x$ and $y$ are,
respectively, the radial and azimuthal coordinates. The shear
parameter $q=1.5$ for the Keplerian rotation considered in this
paper.

The equation of state is
$$
P=(\gamma-1)U,
$$
where $U$ and $\gamma$ are the two-dimensional internal energy density and
adiabatic index, respectively. We will adopt $\gamma=2$ throughout.

The central quantity of this study is the vertical component of
potential vorticity referred to as PV for short:
$$
I\equiv\frac{{\bf \hat{z}}\cdot \nabla\times {\bf u}+(2-q)\Omega}{\Sigma}=\frac{1}{\Sigma}\left(\frac{\partial u_y}{\partial
x}- \frac{\partial u_x}{\partial y}+(2-q)\Omega \right),
$$
where ${\bf \hat{z}}$ is the unit vector in the vertical direction.
In the unperturbed state, where there is only the background
Keplerian shear flow, ${\bf u_0}$, with the constant equilibrium
surface density $\Sigma_0$, PV is equal to
$I_0=(2-q)\Omega/\Sigma_0$. The PV will play an important role in
the subsequent analysis, as it generally characterizes the formation
of coherent structures (vortices) in a disc flow
\citep[e.g.,][]{GL99a,GL99b,JG05,Bo07}. Using equations (1-3), after
some algebra, one can show that the evolution of PV is governed by
the following equation
$$
\left(\frac{\partial}{\partial t}+{\bf u}\cdot\nabla-q\Omega
x\frac{\partial}{\partial y}
\right)I=\frac{1}{\Sigma^3}\left(\frac{\partial \Sigma}{\partial
x}\frac{\partial P}{\partial y}-\frac{\partial \Sigma }{\partial
y}\frac{\partial P}{\partial x}\right).
$$
This equation describes the advection of PV along the trajectories
of fluid elements (Lagrangean derivative inside the brackets on the
left hand side) and its change due to the nonlinear baroclinic term
on the right hand side. In the present case, the pressure and
surface density are not related by any (e.g., isentropic, isothermal
or polytropic) constraint, so this baroclinic term is, in general,
non-zero and, therefore, PV is not conserved. However, in the linear
approximation it vanishes and PV is conserved making it possible to
classify modes into vortical and wave types \citep{MC07}. Also note
that the gravitational potential, as it should be, does not
explicitly enter into this equation; self-gravity only influences PV
evolution through surface density, pressure and velocity fields.

The evolution of internal energy density is governed by the equation
\begin{equation}
\frac{\partial U}{\partial t}+ \nabla\cdot(U{\bf u})-q\Omega x
\frac{\partial U}{\partial y} = -P\nabla\cdot{\bf
u}-\frac{U}{\tau_c},
\end{equation}
where the first term on the right hand side is the compressional
heating term and the second term takes into account cooling of the
disc. Here we assume the cooling time $\tau_c$ to be constant and
choose its value so that the disc does not fragment and achieves a
saturated state, where all quantities fluctuate around constant
average values. Namely, we take $\tau_c=20\Omega^{-1}$, which means
a non-fragmenting disc according to Gammie's (2001) criterion.
\footnote[1]{We also checked that as long as a disc does not
fragment, different values of the cooling time do not qualitatively
change results.} We refer to this quasi-steady state as
gravitoturbulence. \emph{In the present study we concentrate on
examining the specific properties of PV evolution in such a
gravitoturbulent state}. Where fragmentation conditions are
concerned, more realistic cooling laws, with $\tau_c$ being a
function of $\Sigma, U, \Omega$ rather than a constant, are
necessary \citep[e.g.,][]{JG03,Boletal06}. Most simulations of
self-gravitating discs with such a realistic cooling, however,
indicate that in this case cooling is usually not sufficient to
cause fragmentation over most of the disc except possibly for the
outer regions, therefore, a quasi-steady state is more likely (see
review by \citealt{Duetal07}). We do not include any artificial
heating terms in the internal energy equation; heating is solely due
to the compressional term and to shocks captured by means of
artificial (von Neumann-Richtmyer) viscosity in our code.

\subsection{Numerical methods}

Our computational domain in the $(x,y)$ plane is a rectangle
$-L_x/2\leq x \leq L_x/2, -L_y/2\leq y \leq L_y/2$ of size
$L_x\times L_y$ divided into $N_x\times N_y$ grid cells. The
numerical resolution is therefore $\triangle x \times \triangle y =
L_x/N_x \times L_y/N_y$. We will assume that $L_x=L_y\equiv L$ and
$N_x=N_y\equiv N$. For the fiducial model presented below we take
$N=1024$, though we also ran a lower resolution ($N=512$) model that
converges to the fiducial model. In order to integrate the governing
equations (1-5) within this domain, we use a version of the ZEUS
code \citep{SN92}, which is more suited to the shearing sheet
\citep{G01,JG03,JG05}. ZEUS evolves these equations on a staggered
mesh in a time-explicit, operator split, finite-difference fashion.
As mentioned above, the code uses artificial viscosity to capture
shocks.

The transport scheme differs from that of the basic ZEUS algorithm.
To show this, in the left hand side of equations (1-5) we have
explicitly decomposed the advection of physical quantities by the
flow into two parts: the first one (with ${\bf u}\cdot \nabla$)
represents advection by the perturbed velocity and is done using the
van Leer algorithm for calculating fluxes of conserved quantities as
in the original ZEUS algorithm. The second one (with $-q\Omega
x\cdot\partial/\partial y$) describes the advection by the
background Keplerian shear flow ${\bf u_0}$ and is done by means of
the FARGO scheme \citep{M00} with the use of the same van Leer
algorithm for performing the fractional part of the shift; the
integral part of the shift is done by a simple remap. This scheme
has the advantage that it allows larger than standard integration
time steps by removing the background shear flow from the Courant
condition, which severely limits the time step because of large
shear velocities at the $x$-boundaries. Now only the perturbed
velocity ${\bf u}$, rather than the full velocity ${\bf u_0}+{\bf
u}$, enters this condition. Numerical diffusion of the code is also
reduced by this procedure.

As with most works employing shearing sheet approximation, we adopt
periodic boundary conditions in the $y$-direction and shearing
periodic in the $x$-direction, that is, the $x$-boundaries are
initially periodic but as time goes by they shear with respect to
each other becoming again periodic when $t_n=nL_y/(q\Omega
L_x)=n/(q\Omega)$, with $n=1,2,...$. So, for each variable we can
write:
$$
f(x,y,t)=f(x,y+L,t)~~~~~~(y~~boundary)
$$
$$
f(x,y,t)=f(x+L, y-q\Omega Lt, t)~~~~~~(x~~boundary),
$$
where $f\equiv (u_x, u_y, P, \Sigma, U)$. The way of implementing
these shearing sheet boundary conditions is described in detail in \citet{HGB95}.
The shearing in the $x$-boundary conditions,
or shift in the $y$-direction by an amount $-q\Omega Lt$ in the
$''$ghost zones$''$, is done here by means of the FARGO algorithm.

We define the autocorrelation function for PV as
$$
R_I(x,y)=\frac{\Sigma_0^2}{\Omega^2 L_xL_y}\int\delta I(x',y')\delta I(x+x',y+y')dx'dy',
$$
where the integration is over the entire rectangular simulation
domain and $\delta I=I-I_0$. The autocorrelation function
characterizes emerged coherent structures in a flow; its length
scale can be identified with the characteristic scale of such
structures. An analogous function was used by \citet{G01} to analyse
density structures in order to establish locality of angular
momentum transport and by \citet{JG05} to characterize coherent
vortices in the non-self-gravitating shearing sheet.

Our method of solving Poisson's equation is analogous to that used
by \citet{Johetal07} and differs from the original method of
\citet{G01} in the following respect. In \citet{G01}, the surface
density is first linearly interpolated onto a grid of shearing
coordinates $(x, y'\equiv y-q\Omega
x(t-t_p),~t_p=(q\Omega)^{-1}NINT(q\Omega t)) $, where it is exactly
periodic, and then fourier transformed with a standard FFT
technique. After that the fourier transform of the gravitational
potential is found from the Poisson's equation rewritten in
wavenumber $(k_x, k_y)$ plane
$$
\psi(k_x,k_y,t)|_{z=0}=-\frac{2\pi G\Sigma(k_x,k_y,t)}{\sqrt{(k_x+q\Omega k_y
(t-t_p))^2+k_y^2}},
$$
where $\psi(k_x,k_y,t)|_{z=0}$ and $\Sigma(k_x,k_y,t)$ are the
fourier transforms of the gravitational potential and surface
density with the background constant value subtracted. The potential
is then transformed back into the grid of shearing coordinates and
finally linearly interpolated onto the grid of $(x,y)$ coordinates.
We instead fourier transform directly from the $(x,y)$ plane to the
$(k_x, k_y)$ plane and vice versa without doing an intermediate
transformation to the shearing coordinates. However, in this case we
take into account the fact that the radial wavenumber $k_x$ of each
spatial fourier harmonic is no longer constant, but changes with
time as $k_x(t)=k_x(0)+q\Omega k_yt$ in order to remain consistent
with the shearing sheet boundary conditions discussed above. As a
consequence, at each time $t$ in the corresponding computational
domain in the $(k_x, k_y)$ plane there are only fourier harmonics
with wavenumbers $k_{y}=2\pi n_y/L, k_{x}=2\pi n_x/L+q\Omega t'(2\pi
n_y/L)$, where $t'=mod(t,1/(q\Omega|n_y|))$ ($mod(a,b)$ is modulus
after dividing $a$ by $b$) and integer numbers $n_x, n_y$ lie in the
interval $-N/2 \leq n_x, n_y \leq N/2$. So, when doing fourier
(inverse fourier) transform we decompose (sum) into (over) these
wavenumbers. We accordingly modified the standard FFT routine to
perform this decomposition (summation). In some sense it is
equivalent to changing from the $(x,y)$ plane to the $(x,y')$ plane
via fourier transformation, which is more exact than that done by
linear interpolation. \footnote[2]{See website
http://imp.mcmaster.ca/$\sim$colinm/ism/rotfft.html} In order to
make the gravitational force isotropic on small scales, we keep only
harmonics with $k<\pi N/(L\sqrt{2})$, where $k=(k_x^2+k_y^2)^{1/2}$.
We have checked using the linear shearing wave test that the present
version of potential solver gives more accurate results than that
used previously.

This version of the ZEUS code was tested on linear problems
involving shearing (vortical and compressible) waves in the
(non)-self-gravitating shearing sheet \citep{G01,JG05}. The code
results are in an excellent agreement with those of the linear
theory until the decreasing radial wavelength of shearing waves
becomes comparable to the grid cell size. So, we do not give these
basic tests here and refer the interested reader to the above
papers.

\subsection{Total energy equation and $\alpha$ parameter}

In this subsection we derive a relationship between the total energy
and hydrodynamic and gravitational stresses. This relationship is
important as it clearly shows the role of shear in the energy
exchange between the background Keplerian flow and perturbations.
The total energy density of perturbed (from Keplerian shear flow)
gas motion is defined as
$$
E=\frac{1}{2}\Sigma{\bf u}^2+U+\frac{1}{2}(\Sigma-\Sigma_0)\psi,
$$
where the first term is the kinetic energy density, the second term,
as defined above, is the internal energy density and the third term
is the gravitational energy density. Using basic equations (1-5) and
the shearing sheet boundary conditions, the equation for the average
total energy is:
$$
\frac{d\langle E \rangle}{dt}=q\Omega\left\langle\Sigma
u_xu_y+\frac{1}{4\pi G}\int_{-\infty}^{+\infty}\frac{\partial \psi}{\partial x}\frac{\partial \psi}{\partial y}dz\right\rangle-
\frac{\langle U\rangle}{\tau_c},
$$
where the angle brackets denote averages defined as $\langle f
\rangle=\int fdxdy/L_xL_y$, where the integration is done over the
entire simulation domain. The first and second terms inside the
angle brackets on the right hand side are Reynolds and gravitational
stresses respectively. As is clear from this equation, these two
terms are responsible for the energy exchange between the background
Keplerian flow and perturbations, which is in fact possible due to
the differential nature (shear) of Keplerian rotation. In the
absence of background shear ($q=0$) the total energy of
perturbations changes only due to cooling. Incidentally, the sum in
the angle brackets is also proportional to the angular momentum flux
density \citep{G01}. Since the cooling time is not short enough, the
system, as mentioned above, settles into a quasi-steady
gravitoturbulent state in which $d\langle E\rangle/dt=0$ (similarly
for every average variable). Therefore, at these times we can write:
$$
q\Omega\left\langle\Sigma
u_xu_y+\frac{1}{4\pi G}\int_{-\infty}^{+\infty}\frac{\partial \psi}{\partial x}\frac{\partial \psi}{\partial y}dz\right\rangle=
\frac{\langle U\rangle}{\tau_c},
$$
or in terms of the well-known $\alpha$ viscosity parameter
$$
\alpha\equiv \frac{1}{\langle\Sigma c_s^2\rangle}\left\langle\Sigma
u_xu_y+\frac{1}{4\pi G}\int_{-\infty}^{+\infty}\frac{\partial \psi}{\partial x}\frac{\partial \psi}{\partial y}dz\right\rangle
$$
giving
\begin{equation}
~~~~~~~~~~~~~~~~~~~~~~~~~~~ \alpha=\frac{1}{q\gamma(\gamma-1)\Omega\tau_c} ,
\end{equation}
where $c_s$ is the adiabatic sound speed defined as $c_{s}^2=\gamma
P/\Sigma=\gamma (\gamma -1)U/\Sigma$. This relation is very
important, as it shows that \emph{if} a quasi-steady state can be
achieved, then angular momentum transport must be determined solely
by the cooling time for a given rotation rate and, therefore, should
be constant with time. (The same formula for $\alpha$ was derived by
\citealt{G01} in a different way.) It is the local analog of the
result that the surface brightness of the disc does not depend on
the details of the angular momentum transport mechanism. Again, the
basic premise here is the possibility of the existence of a
quasi-steady gravitoturbulent state, which is able to feed itself,
or be self-sustained, at the expense of background shear flow energy
extracted by Reynolds and gravitational stresses. In the presence of
disc self-gravity and cooling this state is easily reached because
of the self-regulation mechanism \citep{BL01}. However, in thin
two-dimensional models of non-self-gravitating neutral discs steady
outward angular momentum transport is hard to achieve. In this case
typically the disc is unstable to vortex formation and develops
well-organised vortices, which although able to transport angular
momentum outwards mostly via generating compressible motions
(shocks), are still slowly decaying on the time scale of several
hundred orbital periods \citep[e.g.,][]{JG05,SSG06}. Thus, the role
of self-gravity and cooling is crucial for the maintenance of
gravitoturbulence.

Before proceeding to the main analysis, we introduce non-dimensional
variables. As mentioned above, in the unperturbed state the
background surface density $\Sigma_0$, pressure $P_0$ and internal
energy $U_0$ are all spatially constant. In the unperturbed state
the sound speed is $c_{s0}^2=\gamma P_0/\Sigma_0=\gamma
(\gamma-1)U_0/\Sigma_0$. We switch to non-dimensional variables:
$t\rightarrow \Omega t, (x,y)\rightarrow (x\Omega/c_{s0},
y\Omega/c_{s0}), (k_x,k_y)\rightarrow (k_x c_{s0}/\Omega, k_y
c_{s0}/\Omega)$, correspondingly the computational domain size $L
\rightarrow L\Omega/c_{s0}, (u_x, u_y, c_s)\rightarrow (u_x/c_{s0},
u_y/c_{s0}, c_s/c_{s0}), \Sigma \rightarrow \Sigma/\Sigma_0, (P, U,
E)\rightarrow (P/c_{s0}^2\Sigma_0, U/c_{s0}^2\Sigma_0,
E/c_{s0}^2\Sigma_0), \psi\rightarrow \psi/c_{s0}^2, I\rightarrow
I\Sigma_0/\Omega$. Note that $c_{s0}/\Omega$ is actually the scale
height of the disc. Therefore, distances are normalised by the disc
scale height. The local Mach number is defined as
$M=\sqrt{u_x^2+u_y^2}/c_s$. The Toomre parameter, a measure of the
self-gravity of a disc, is $Q=c_s\Omega/\pi G\Sigma$ \citep{To64}.
These non-dimensional variables are used from now on throughout the
paper. In the simulations presented below, we start with
$Q=Q_0=c_{s0}\Omega/\pi G\Sigma_0=1.5$.

\begin{figure}
\includegraphics[width=\columnwidth]{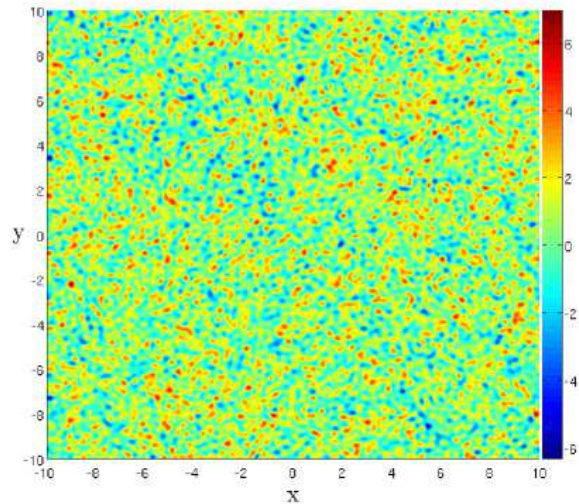}
\caption{Initial (at $t=0$) distribution of PV in the real plane
corresponding to a Kolmogorov spectrum in wavenumber plane for the
fiducial model with $L=20$. Both negative (blue and light blue) and
positive (yellow and red) values of PV are present.}
\end{figure}

\subsection{Initial conditions}

The initial conditions consist of random $u_x$ and $u_y$
perturbations superimposed on the mean Keplerian shear flow. The
surface density and internal energy are not perturbed initially and
thus are spatially uniform with values $1$ and $1/\gamma(\gamma-1)$
respectively. To generate these initial conditions, for each
velocity component at each point in the $(k_x,k_y)$ plane we create
a Gaussian random field with standard deviation, or amplitude of
power spectrum, given by the Kolmogorov power law
$|u_{x,y}(k_x,k_y)|^2\sim k^{-8/3}$, $k=\sqrt{k_x^2+k_y^2}$, in the
range $k_{min}\leq k \leq k_{max}$. The limits of this range are
$k_{min}=2\pi/L$ and $k_{max}=(N/n_g)k_{min}$, where $n_g$ is the
number of grid cells contained within the smallest wavelength
$2\pi/k_{max}$ for which we choose $n_g=16$ when $N=1024$. Outside
this wavenumber range the power spectrum is set to zero. This random
field is then transformed back into the real $(x,y)$ plane. For the
amplitude of velocity perturbations $\sigma = \langle {\bf
u}^2\rangle^{1/2}$, we initially take $\sigma=0.6$. Figure~1 shows
these initial conditions in terms of PV in the real plane. From this
figure we can see that in the real plane the initial PV field has a
chaotic character with equal contributions from negative and
positive values. Although we start with a Kolmogorov power spectrum,
a general dynamical picture emerging in the quasi-steady state does
not depend strongly on the initial conditions. These random velocity
perturbations are meant to mimic the turbulent state in a disc
resulting from the collapse of a molecular cloud \citep{GL00}.

Note that we do not impose any constraints on the initial fields
$u_x$ and $u_y$, such as the requirement of incompressibility
($\nabla\cdot{\bf u}=0$) adopted by \citet{JG05} and \citet{SSG06}
to pick out only vortical perturbations; so these two velocity
components are not correlated. We found that the mere presence of
negative PV values in initial conditions is sufficient for the
development of vortices, even though they may also contain some
fraction of wave perturbations. Linear analysis \citep{MC07} shows
that in shear flows vortical perturbations become divergent if they
are initially non-divergent and wave perturbations become rotational
(with non-zero velocity curl) due to the background velocity shear.
But wave perturbations never acquire PV if it is not present
initially (of course, here we assume that there are no baroclinic
terms in the PV equation in the linear approximation, or
equivalently unperturbed surface density and pressure are uniform).
Thus, truly vortical perturbations should be characterized by
non-zero PV even though they may develop divergence in the course of
evolution. It is the PV that determines the vortex formation and it
must be present initially. By choosing initial perturbed velocity
field with non-zero PV, we ensure the presence of vortical
perturbations with a possible mix of wave perturbations with zero
PV, which do not affect the vortex formation process. In other
wards, the class of initial conditions capable of causing vortex
formation is broader; they must be characterized only by non-zero
negative PV and may not necessarily satisfy the incompressibility
requirement, that is, they can comprise a fraction of wave
perturbations as well. This alleviates the problem of vorticity
injection. Initial conditions resulting from the collapse of a
molecular cloud into a disc, where PV is everywhere zero would be
somewhat contrived and therefore less likely.

\section{nonlinear evolution}

\begin{figure}
\includegraphics[width=\columnwidth]{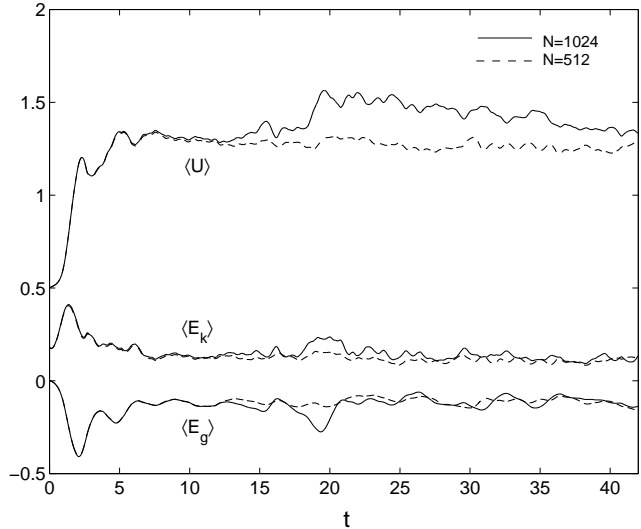}
\caption{Evolution of the average kinetic, internal and
gravitational energies. During the burst phase, they undergo rapid
(swing) amplification until $t\approx 2-2.5$ and then remain on
average constant in the quasi-steady gravitoturbulent phase, which
sets in at about $t=5$. The evolution of the same quantities for a
lower resolution ($N=512$) run are similar.}
\end{figure}
\begin{figure}
\includegraphics[width=\columnwidth]{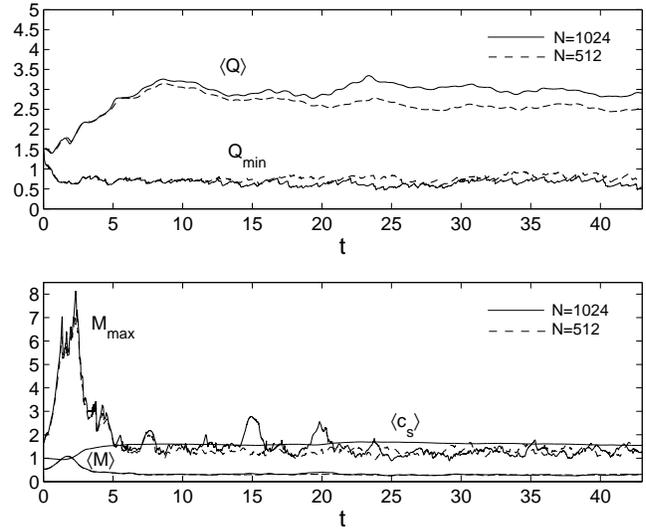}
\caption{Evolution of $\langle Q\rangle$ and minimum $Q_{min}$
(upper plot) and the average Mach number $\langle M\rangle$, maximum
$M_{max}$ and $\langle c_s\rangle$ (lower plot). $\langle Q\rangle$
initially rises due to strong shock heating in the burst phase, but
then levels off at about 3 in the gravitoturbulent phase. $Q_{min}$
remains constant and low 0.6-0.7 implying that there are always
unstable regions associated with negative PV, or vortices. $\langle
M\rangle$ initially rises but then settles down to smaller values
0.3, so that gas motion is on average subsonic. However, $M_{max}$
remains of order unity showing that there is still some sonic
motion. The average sound speed $\langle c_s\rangle$ also reaches a
constant value in the gravitoturbulent phase.}
\end{figure}

In this section we present the nonlinear evolution of the fiducial
model with size $L=20$ starting with the initial conditions
described in the previous section. We describe in detail the
development of structures in the PV, surface density and
pressure/internal energy fields. The size $L=20$ corresponds to a
minimum wavenumber $k_{y~min}=0.314$ in the computation domain, for
which linear swing growth is largest as a function of $k_y$ at
$Q=1.5$. Thus, we ensure that from the outset our computational
domain comprises those scales for which self-gravity is important.
In the presence of both strong Keplerian shear and self-gravity the
main mechanism responsible for the growth of initial velocity
perturbations is swing amplification instead of pure Jeans
instability. Swing amplification has a transient exponential nature
because of the $''$drift$''$ of radial wavenumbers of spatial
fourier harmonics of perturbations through unstable regions in the
$(k_x,k_y)$ plane that are brought about by the combined action of
shear and self-gravity \citep{To81,KO01,MC07}. As a result, the
average values of various quantities undergo rapid amplification
over a relatively short time interval (Figs. 2 and 3; see also
\citealt{G01}. The evolution of the same quantities for a lower
resolution $N=512$ run are almost similar. The small difference
between these two runs may be attributed to the specificity of PV,
surface density and internal energy fields for different
resolutions, see also \citealt{SSG06}). We call this interval the
burst phase. During swing amplification, the initial velocity
perturbations induce strong surface density perturbations in the
form of shocks with density contrast by a factor of about $100$.
These shocks have trailing orientation, because swing amplification
always tends to produce trailing structures. The gas motion is also
supersonic at this stage. It can be seen from Fig.~3 that the
average and maximum Mach numbers reach largest values $1.1$ and
$7-8$ respectively. Shock regions are the sources of intense heating
of the gas together with compressional heating (the first term on
the right hand side of equation 5) and therefore the internal energy
undergoes jumps in the shocks. Eventually strong shocks heat the
disc up, but cooling compensates for heating, so that at about $t=5$
a quasi-steady gravitoturbulent state is reached, where the average
kinetic, internal and gravitational energies as well as $\langle
Q\rangle,~Q_{min},$ the average Mach number $\langle M\rangle,$ the
maximum Mach number $M_{max}$ and $\langle c_s\rangle$ fluctuate
around constant values (Figs.~2 and 3). At these times, the angular
momentum transport parameter $\alpha$ is given by equation (6) and
is constant. Now $\langle M\rangle$ fluctuates around 0.3, so that
on average the gas motion is subsonic, however, $M_{max}$ fluctuates
around 1.24 meaning that the motion is still sonic mostly in the
vicinity of shocks. $\langle Q\rangle$ fluctuates around 3, but we
will see below that the $Q(x,y,t)$ field is very inhomogeneous
containing values as small as 0.6-0.7 associated with negative PV
regions (vortices).

\begin{figure*}
\includegraphics[width=0.32\textwidth, height=0.28\textwidth]{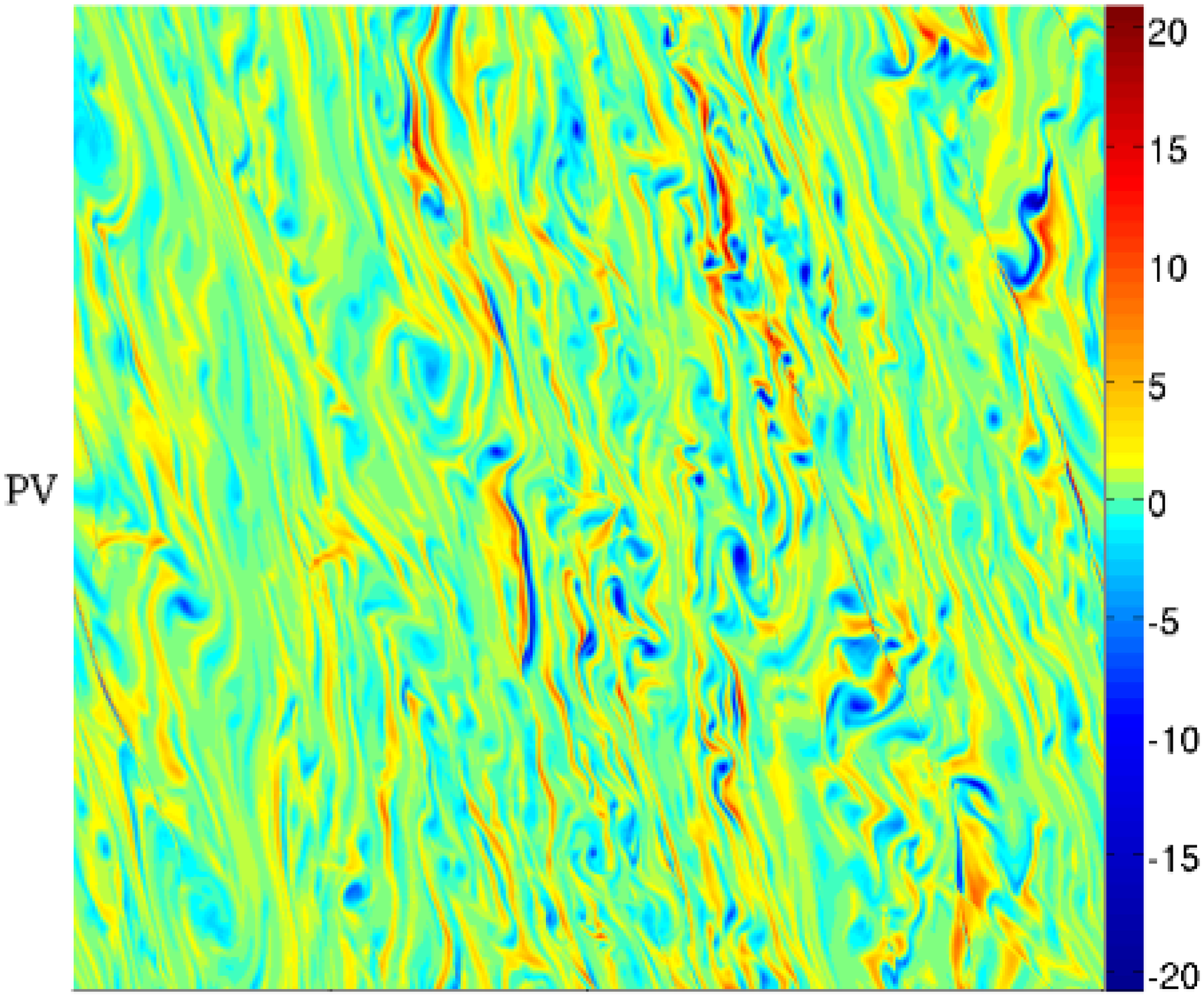}
\includegraphics[width=0.32\textwidth, height=0.28\textwidth]{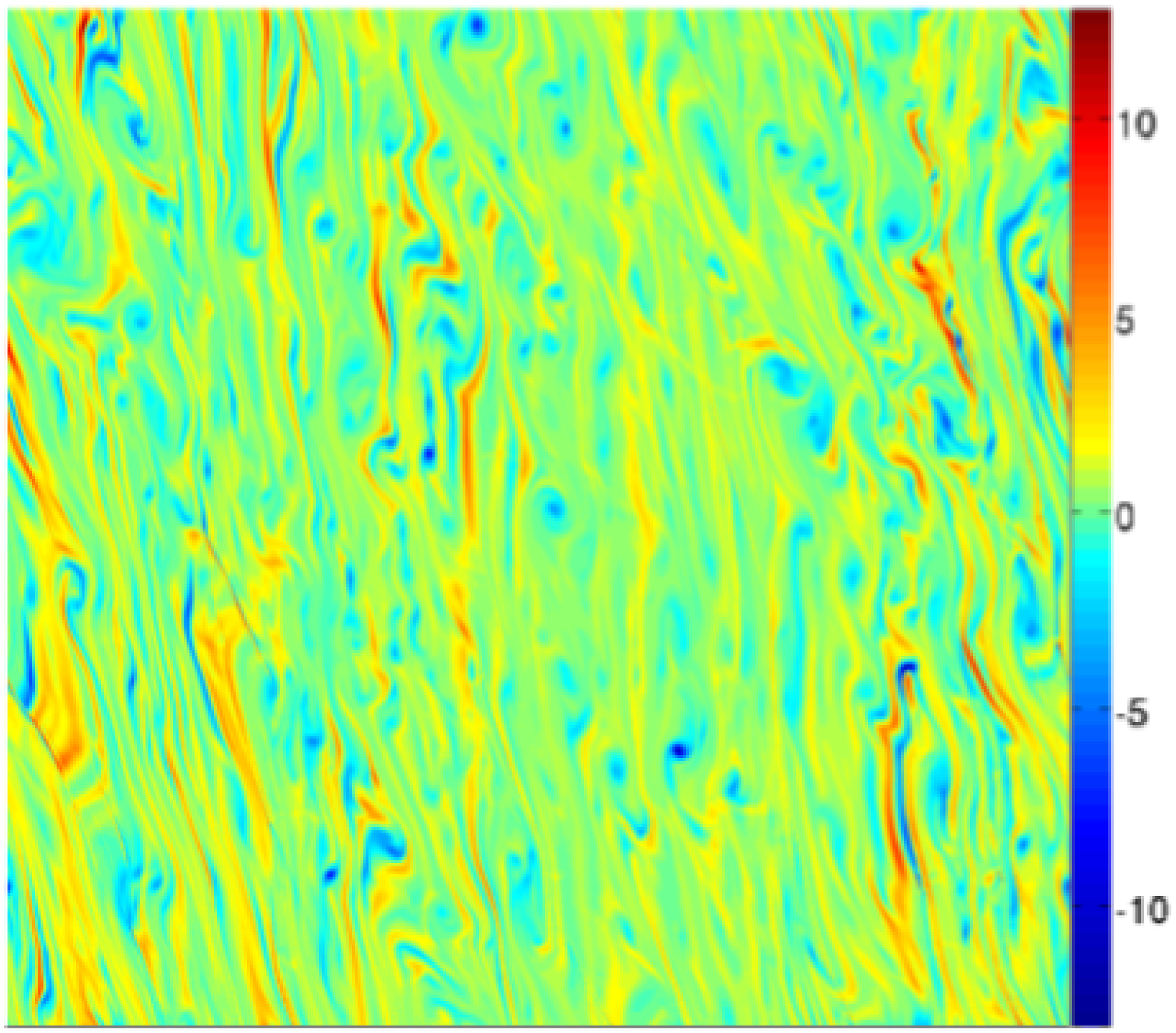}
\includegraphics[width=0.32\textwidth, height=0.28\textwidth]{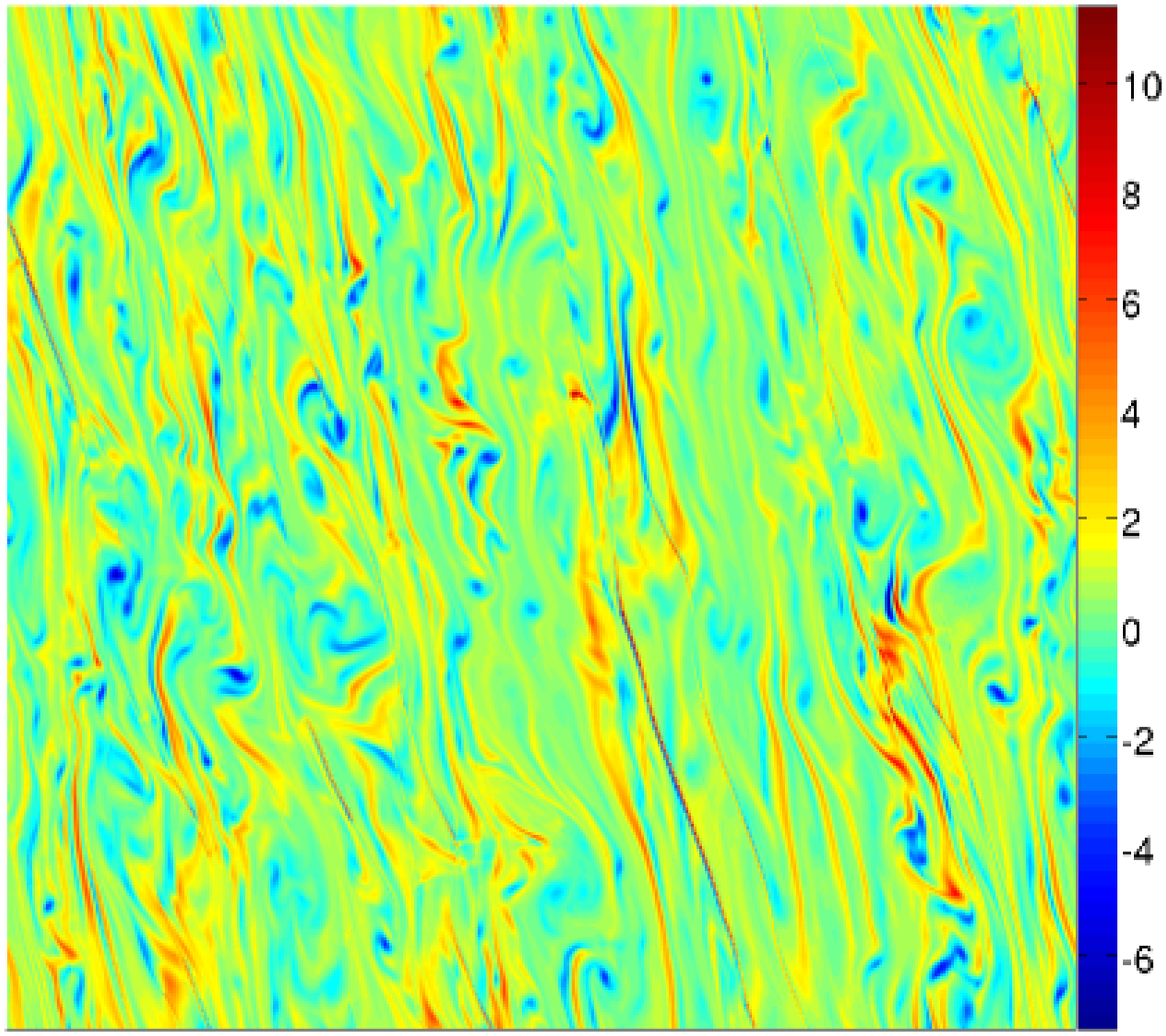}
\includegraphics[width=0.32\textwidth, height=0.28\textwidth]{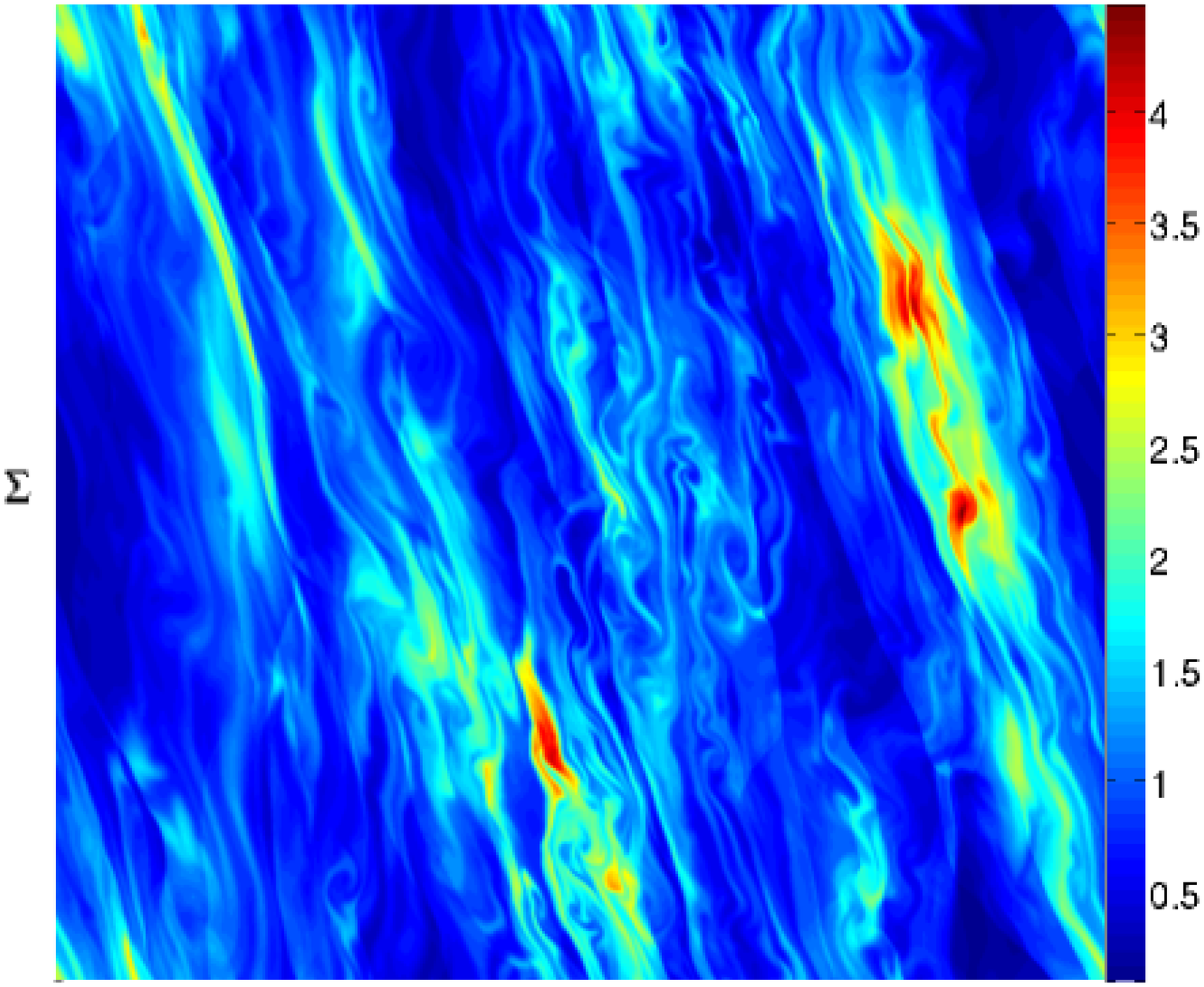}
\includegraphics[width=0.32\textwidth, height=0.28\textwidth]{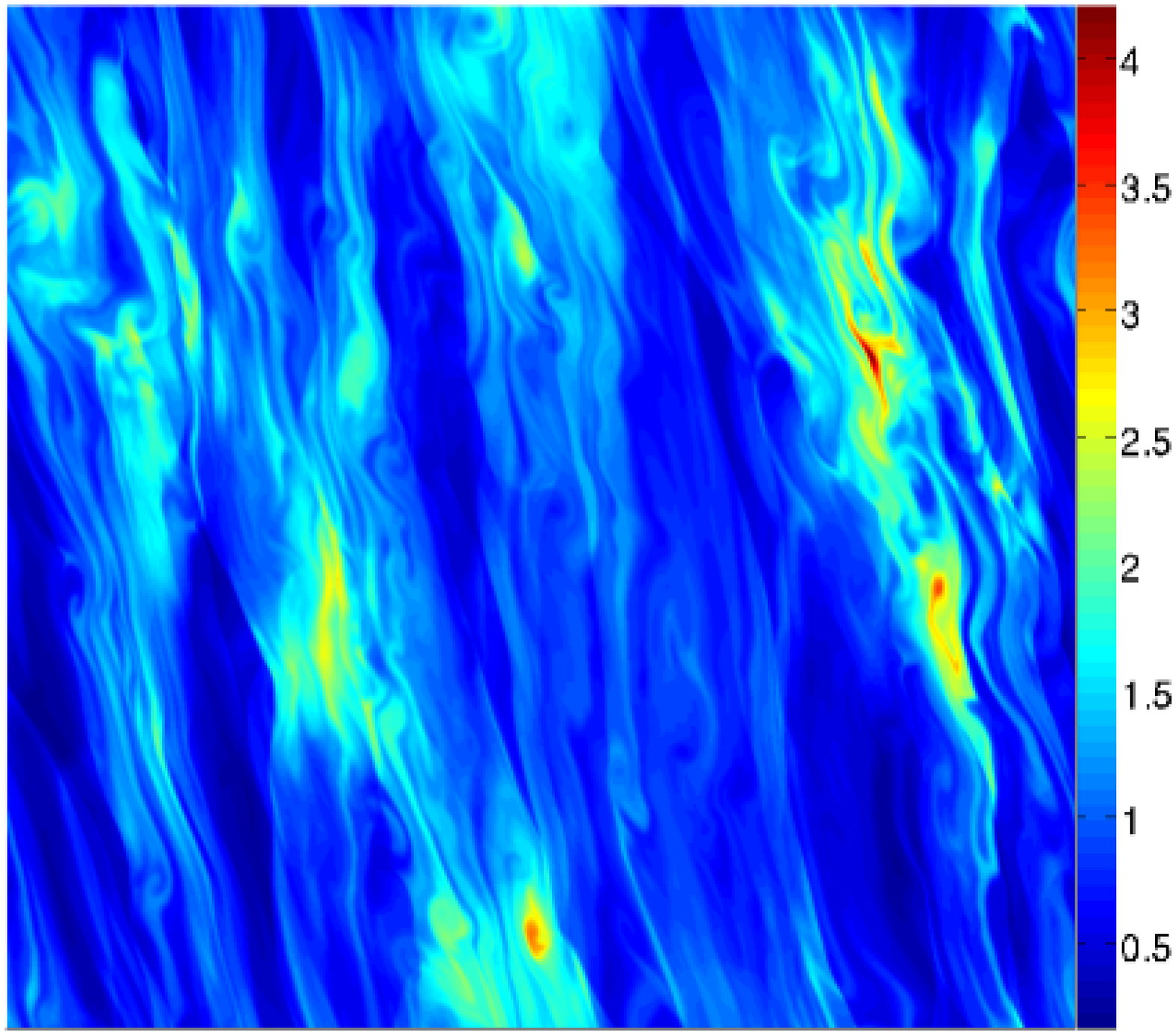}
\includegraphics[width=0.32\textwidth, height=0.28\textwidth]{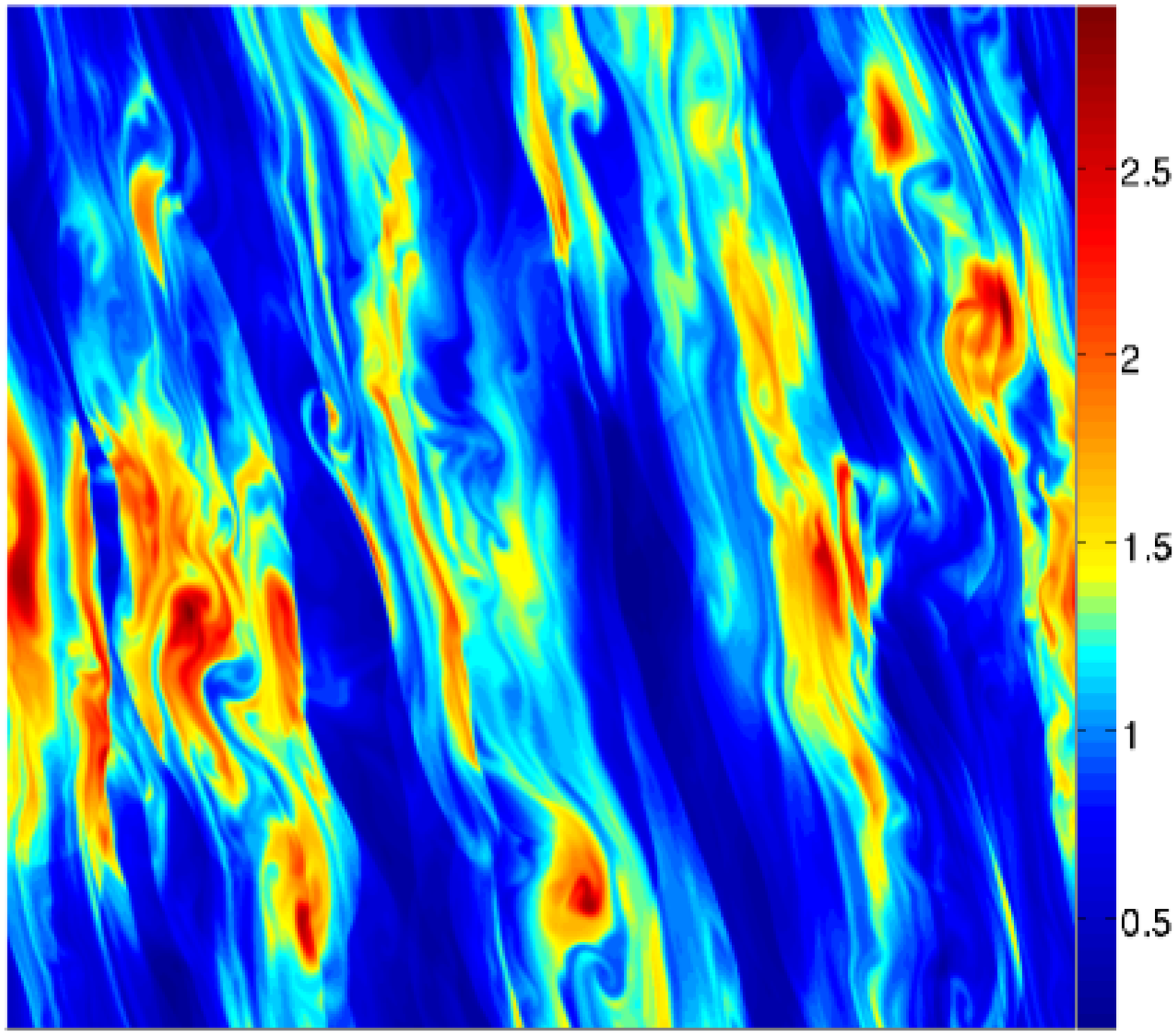}
\includegraphics[width=0.32\textwidth, height=0.28\textwidth]{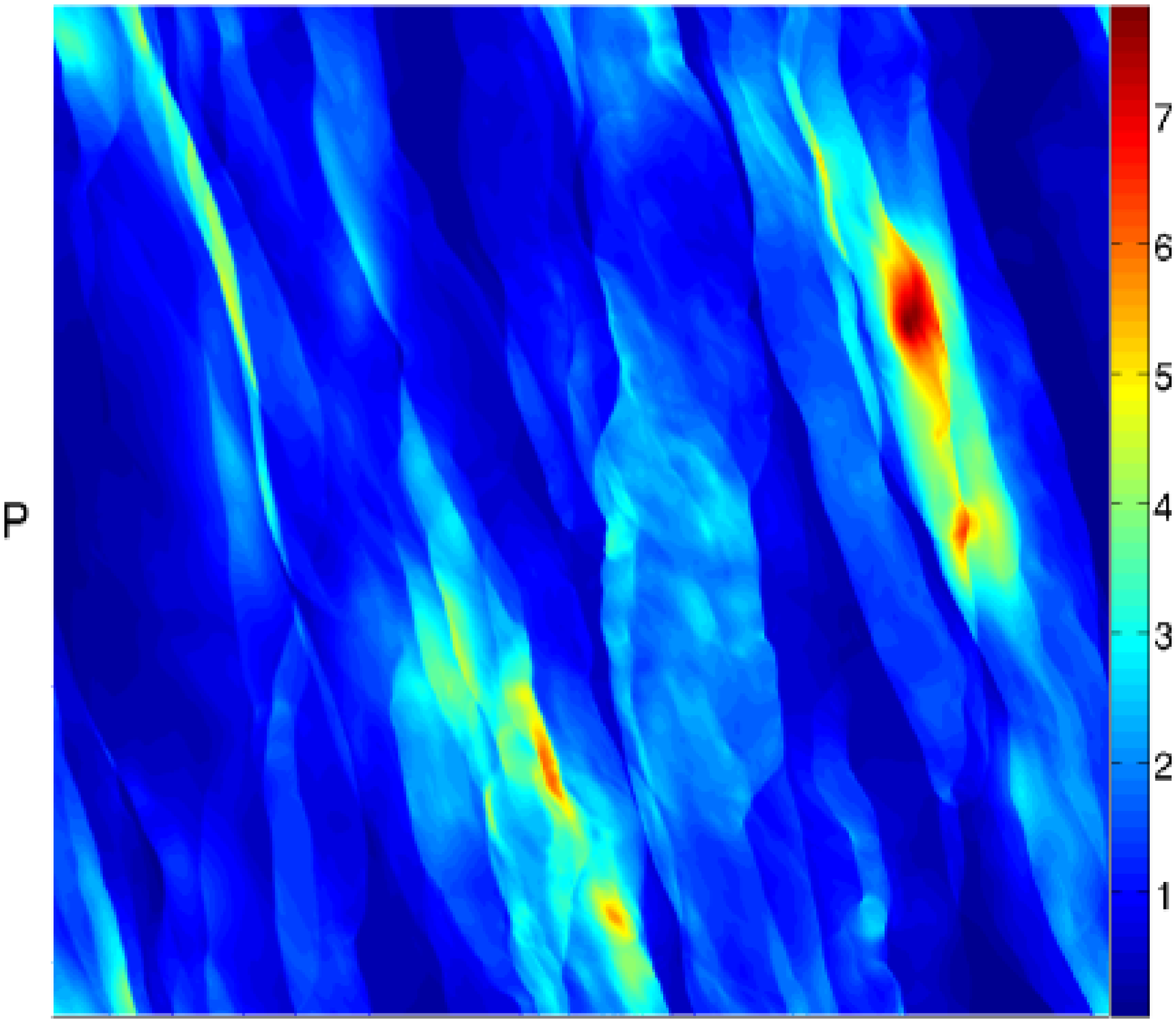}
\includegraphics[width=0.32\textwidth, height=0.28\textwidth]{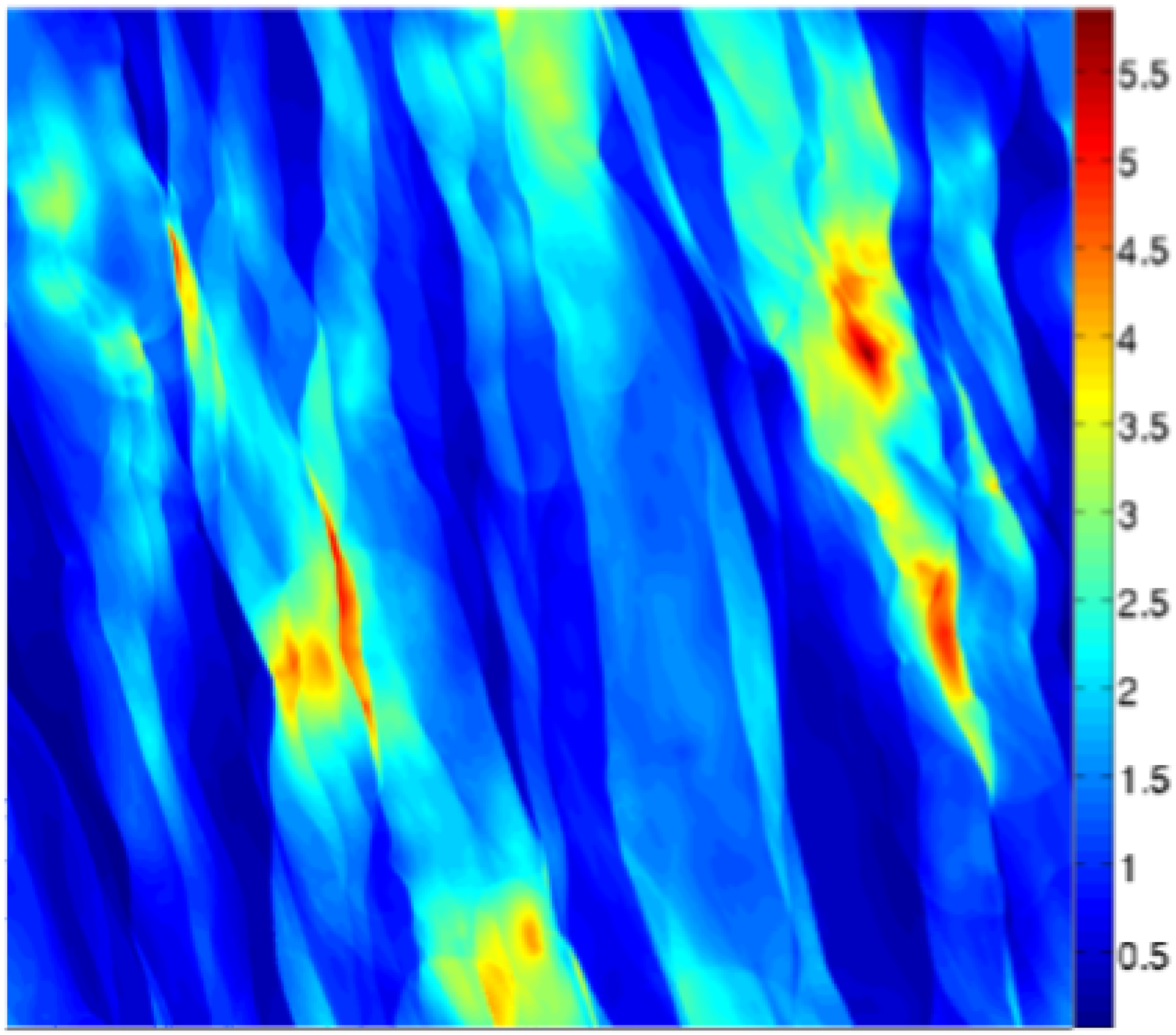}
\includegraphics[width=0.32\textwidth, height=0.28\textwidth]{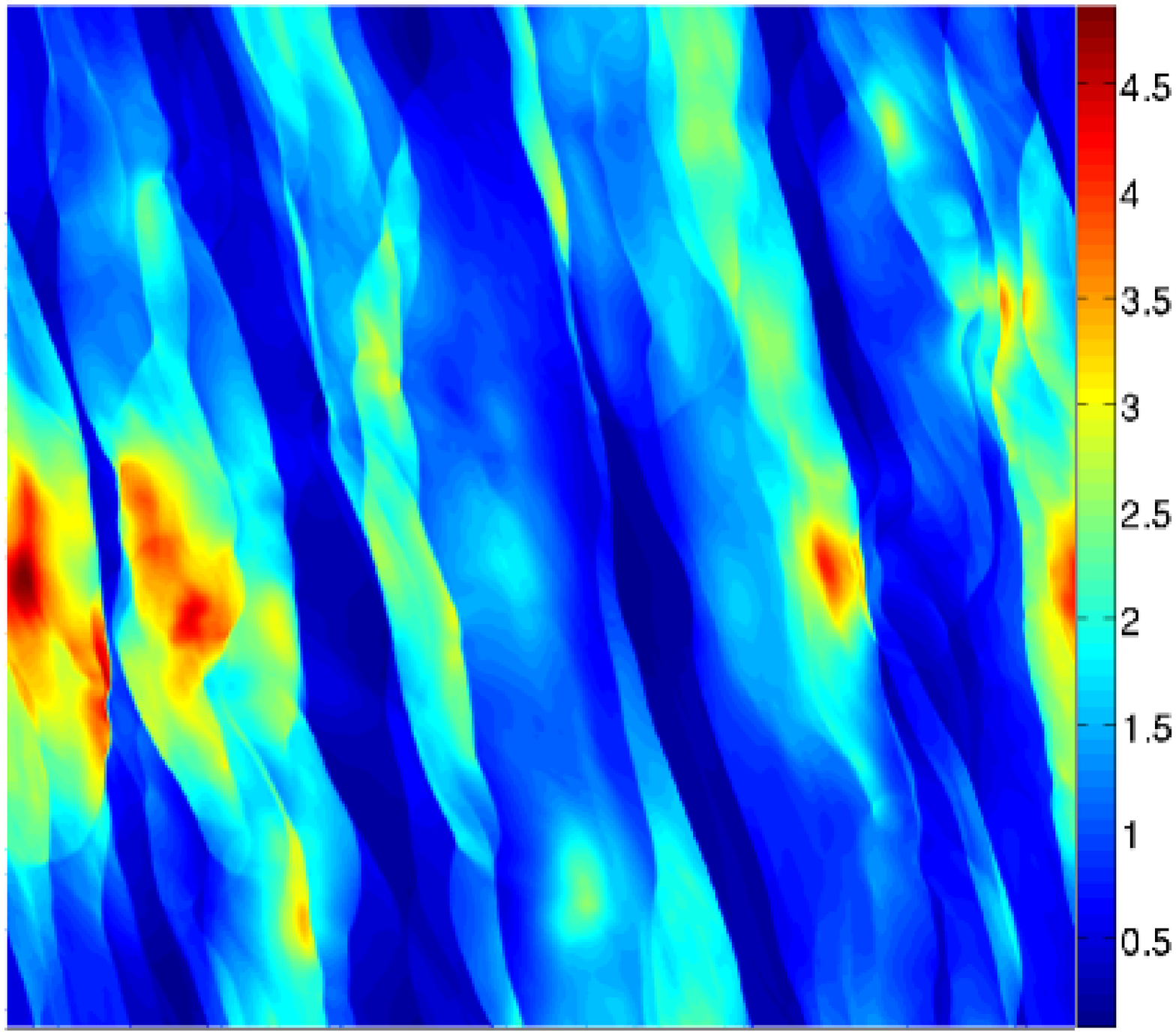}
\includegraphics[width=0.32\textwidth, height=0.28\textwidth]{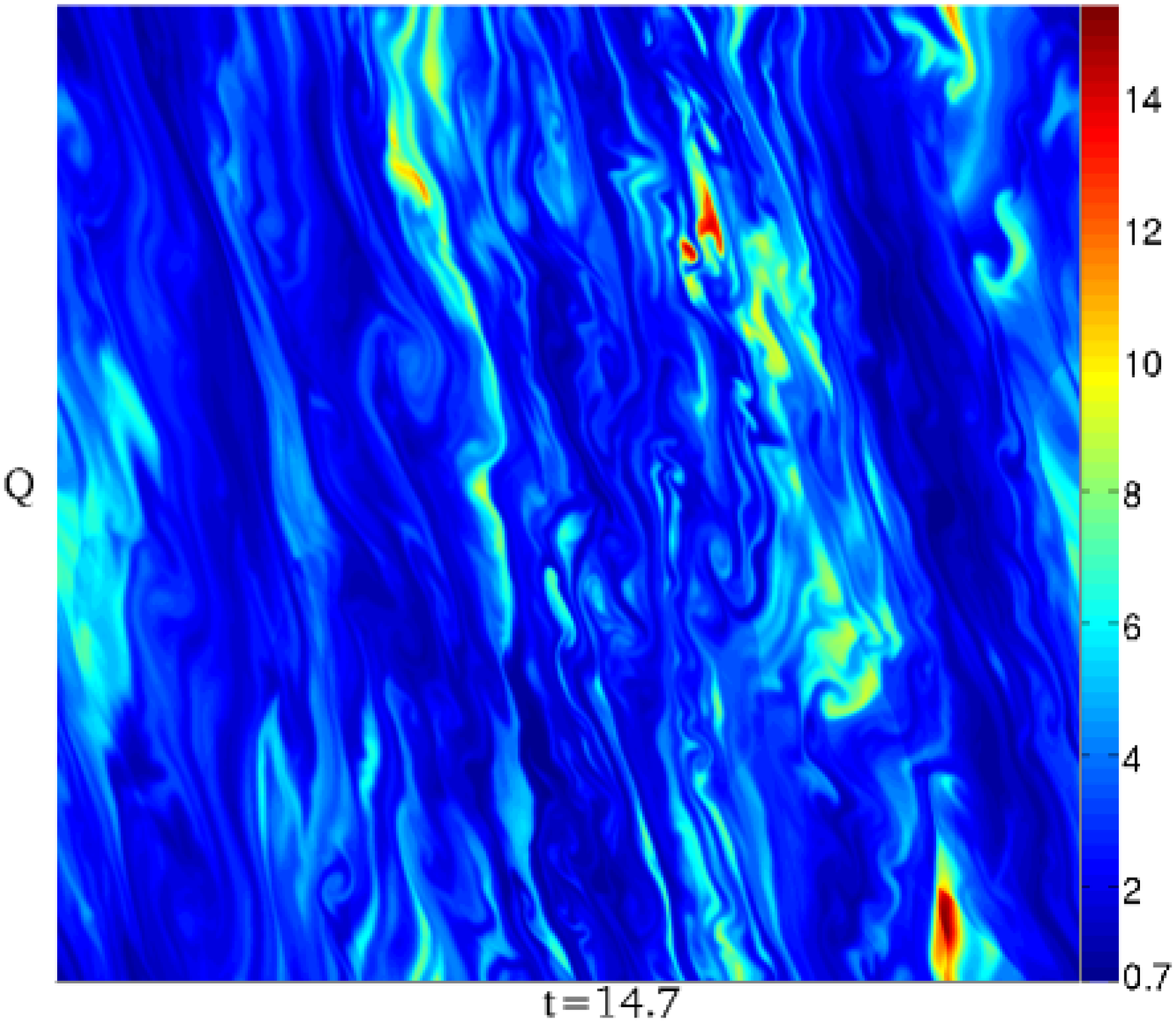}
\includegraphics[width=0.32\textwidth, height=0.28\textwidth]{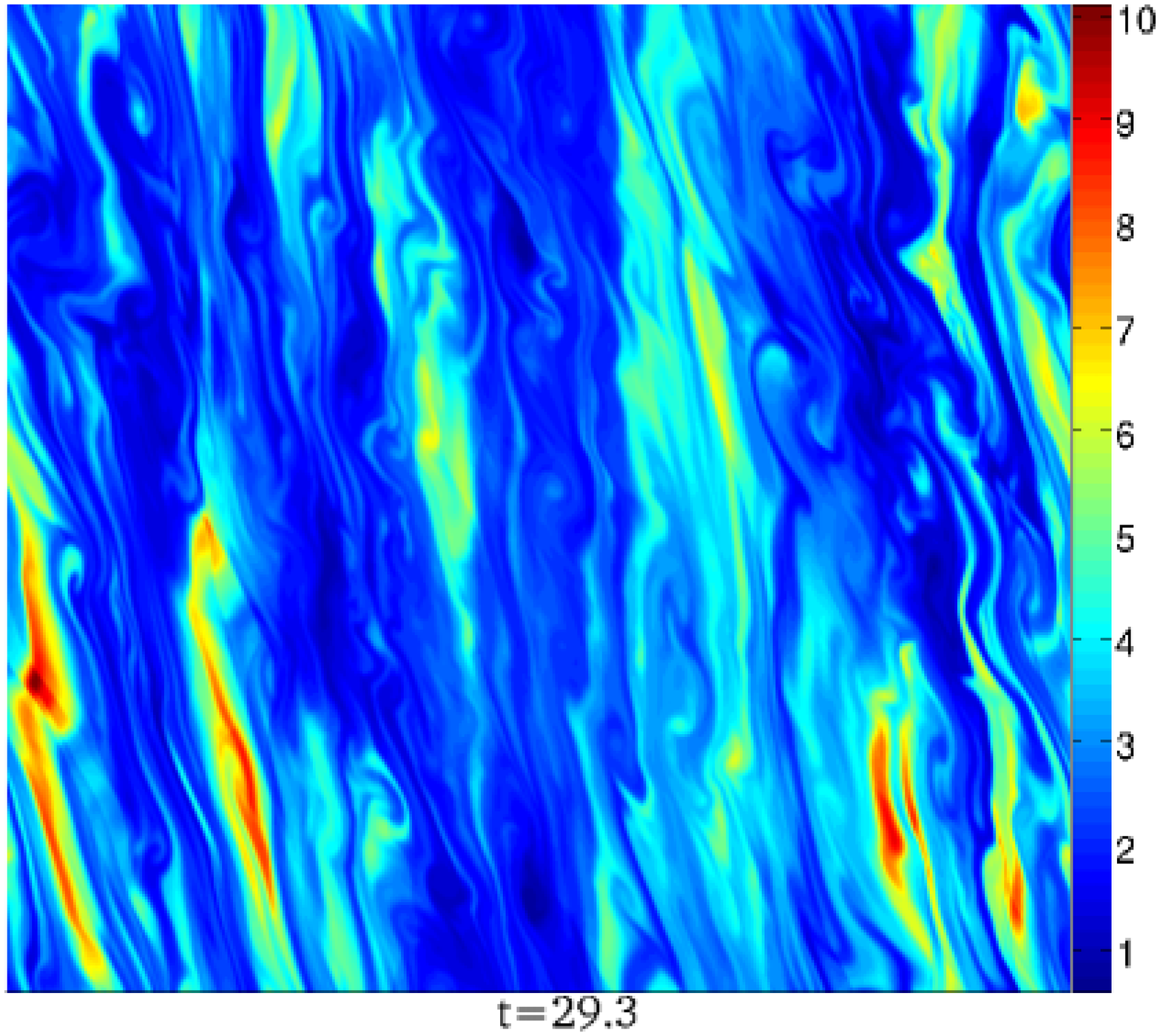}
\includegraphics[width=0.32\textwidth, height=0.28\textwidth]{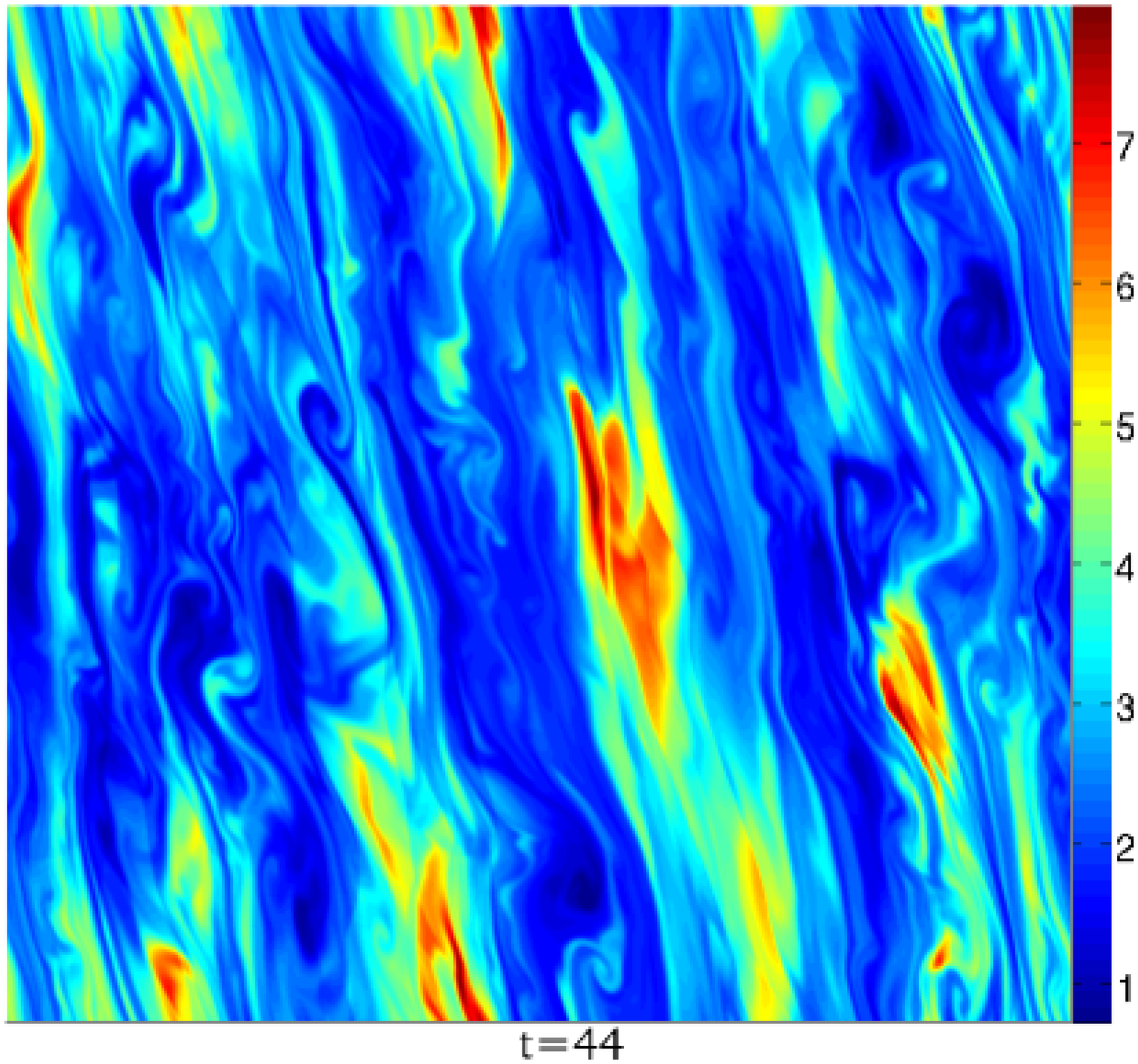}
\caption{Evolution of PV, surface density, pressure/internal energy
and $Q$ parameter (from upper to lower rows respectively) for the
fiducial model in the gravitoturbulent state. Coordinate axes are
same as in Fig.~1. Snapshots are at times $t=14.7, 29.3, 44$. The PV
field consists of vortices with different sizes and strengths (blue
and light blue regions). Vortices with sizes smaller than the local
Jeans scale have higher (by absolute value) negative PV centres
(blue dots). They also emit density waves during evolution that turn
into shocks. In the surface density field, these vortices correspond
to underdense central regions surrounded by overdense regions
marking density wave emission places that are characterized by
smaller (by absolute value) than the central values of PV. In the
$Q$ field, the underdense and overdense regions give rise to high
and low $Q$ values respectively, but the influence of self-gravity
on such small scale vortices is weak. The other type of vortices are
somewhat larger having scales comparable to the local Jeans scale.
They are more diffuse with no clear shape (light blue regions in PV
field) and are characterized by smaller (by absolute value) negative
PV and stronger overdense regions (yellow and red in the surface
density field) with even smaller $Q$ than those in the previous
case. For this reason, such vortices are in the process of being
sheared and destroyed. In the pressure field, shocks and only the
overpressure regions corresponding to stronger overdensities are
noticeable (yellow and red). Exact values of these four quantities
can be inferred from colour bars.}
\end{figure*}

In Fig.~4 we trace the parallel evolution of PV, surface density,
pressure (the same as internal energy for $\gamma=2$) and $Q$ fields
to identify correlations between them and to compare them with those
in non-self-gravitating discs. We emphasize from the beginning that
the evolution is not as smooth and regular as it is in the
non-self-gravitating case. As we will see below, in self-gravitating
discs vortices are transient structures undergoing recurring phases
of formation, growth and destruction, whereas in
non-self-gravitating discs small scale vortices once formed,
gradually merge into larger vortices.

During the burst phase, initially small scale positive and negative
PV regions get strongly sheared into strips, but negative PV
(anticyclonic) regions start to wrap up into vortex-like structures
towards the end of the burst phase due to nonlinear Kelvin-Helmholtz
instability \citep{L07}. The positive PV regions remain sheared into
strips showing no signs of vortex formation during the entire course
of evolution. Thus, only anticyclonic regions are able to survive in
shear flows by taking the form of vortices. As mentioned in the
Introduction, this is a quite general result confirmed in many
simulations of astrophysical as well as purely hydrodynamical
contexts. As evident from Fig.~4, in the subsequent quasi-steady
state the overall dynamical picture of PV evolution is still
irregular and varying on the dynamical timescale. For this reason,
throughout the paper we use the term $''$vortex$''$ in a broader
sense meaning generally a negative PV region and not only a vortex
with a well-defined shape commonly occurring in non-self-gravitating
discs. As demonstrated below, in a quasi-steady state these negative
PV regions are associated with underdense/underpressure and
overdense/overpressure regions superimposed on the shocks that may
be important for trapping dust particles.

Small scale anticyclonic vortices initiated at the end of the burst
phase continue to grow further in size in the quasi-steady saturated
state which is reached, as mentioned, at about $t=5$. Figure~4
presents a typical dynamical picture of such a gravitoturbulent
state, which is quasi-stationary and remains unchanged (on average)
with time during the entire course of evolution. Vortices with a
large (by absolute value) negative central value of PV (blue dots in
the PV field) have more or less vortex-like shape and correspond to
the central underdense regions surrounded by the higher density
regions. Due to the background Keplerian shear, the total motion in
the vicinity of such negative PV regions is a complex mixture of
vortical and compressive (wave) motions giving rise to this type of
density structure. In these enhanced density regions, PV is smaller
by absolute value than that in the vortex centre. Thus, generally in
shear flows the coupling with wave motions is typical of vortex
dynamics irrespective of the disc being self-gravitating or not.
These compressive motions turn into shocks afterwards that are also
seen in the surface density and pressure fields (more precisely, the
shock structure in these fields is a net result of nonlinear
superposition/mixture of shock waves emitted by such individual
vortices). So, classifying perturbations into purely vortical --
having only zero divergence and non-zero curl -- and compressive --
having only non-zero divergence and zero curl -- is not quite
appropriate, because due to background shear, divergence and curl
become linked and change with time during the evolution that, in
turn, causes mixing of these two perturbation types \citep{MC07}.
With regard to angular momentum transport, it would be more exact to
attribute it to the overall gas motion with non-zero PV and
divergence rather than to compressive (wave) or to vortical
perturbations only.

A similar dynamical picture is also observed in the
evolution/adjustment of a single vortex in a compressible
non-self-gravitating Keplerian disc \citep{Bo07}. The vortex
initially produces saddle-like density structure, much like observed
in our simulations, with an underdense region at the location of the
vortex centre surrounded by an overdense region with accompanying
shock waves .

With time the effect of self-gravity on each vortex comes into play
because of the growing length scale of vortices and at the same time
of the increasing surface density/overdensity associated with them.
The vortex growth in size is a consequence of inverse -- towards
larger scales -- cascade of energy characteristic of 2D flows as
clearly demonstrated in simulations \citep[e.g.,][]{GL99a}. The
amplification of the vortex surface density is again due to the
swing mechanism. For the same reason that in self-gravitating discs
perturbations experience considerably larger swing growth than in
non-self-gravitating discs, the traces of vortices in the surface
density field -- the overdense regions around underdense regions --
become more and more noticeable on the background of density
variations related to shocks. From Fig.~4 it is seen that
overdensities are also characterized by small $Q$ values. In the
pressure/internal energy field we see shocks and only overpressure
regions. The stronger overdensities (yellow and red in Fig.~4)
associated with the final stage of vortex life (see below) clearly
correlate with these overpressure regions. However, it is still hard
to find as good a correspondence/similar features between PV and
pressure/internal energy fields as it is for the surface density
field. Nevertheless, in self-gravitating discs one can identify
distinct features in the surface density and pressure/internal
energy fields connected with vortices. By contrast, in analogous
non-self-gravitating isothermal simulations the situation is
different. As mentioned above, similar density structures, i.e.,
with overdense and underdense regions, were also observed in
numerical simulations of a single vortex before its settlement into
an equilibrium configuration, where the anticyclonic vortex gives
rise only to a single overdense region \citep{Bo07}. However, other
simulations of vortex formation from an initial random distribution
of PV \citep{JG05,SSG06}, which are not limited to only a single
vortex, did not find overdense regions easily identifiable with
individual vortices in the surface density field; there were only
variations in the surface density due to the shocks shed by these
vortices (see also Fig.~5). In our case instead, swing amplification
due to self-gravity makes overdense and underdense regions clearly
noticeable on the background of shocks.

\begin{figure}
\centering\includegraphics[width=\columnwidth]{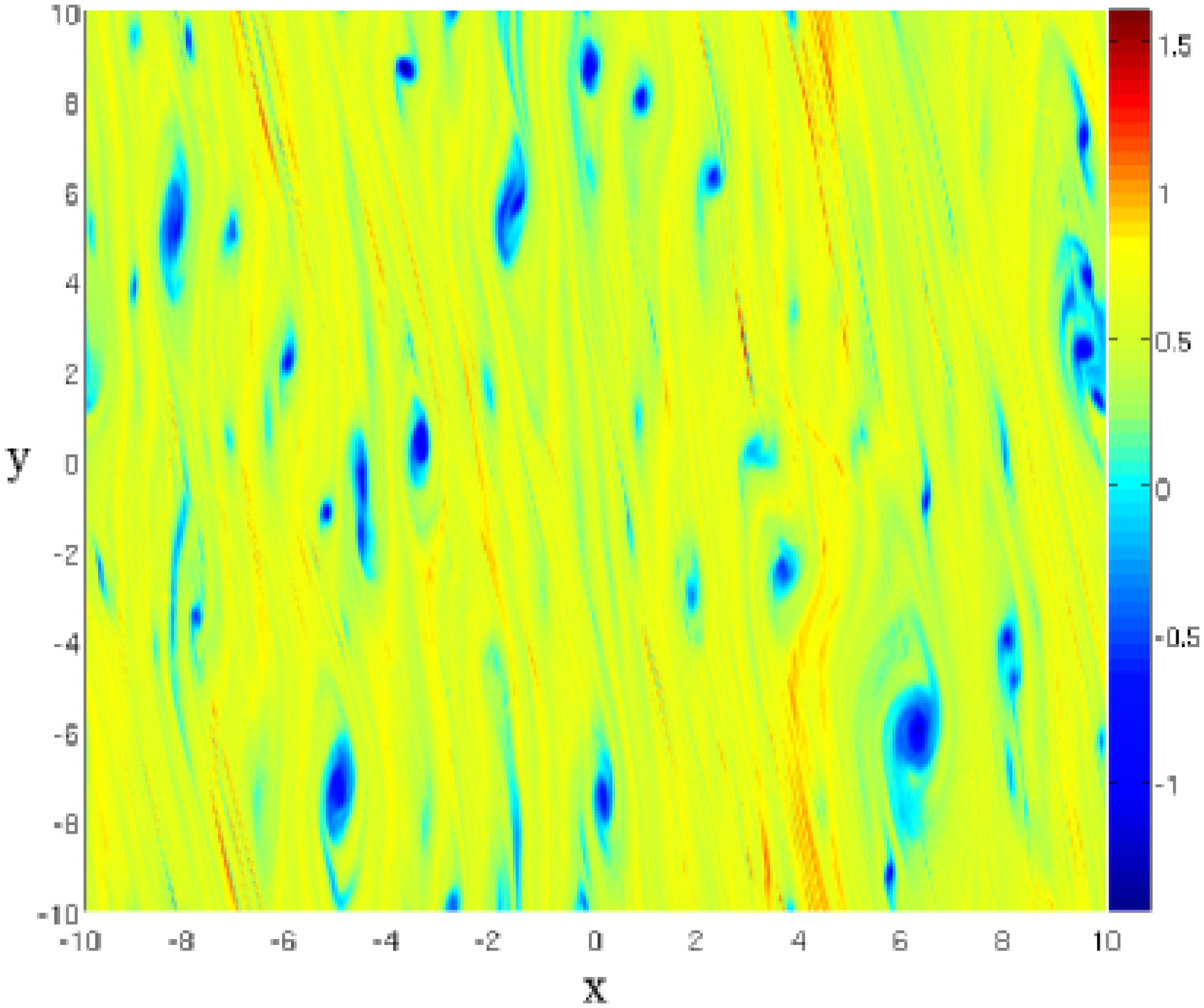}
\centering\includegraphics[width=\columnwidth]{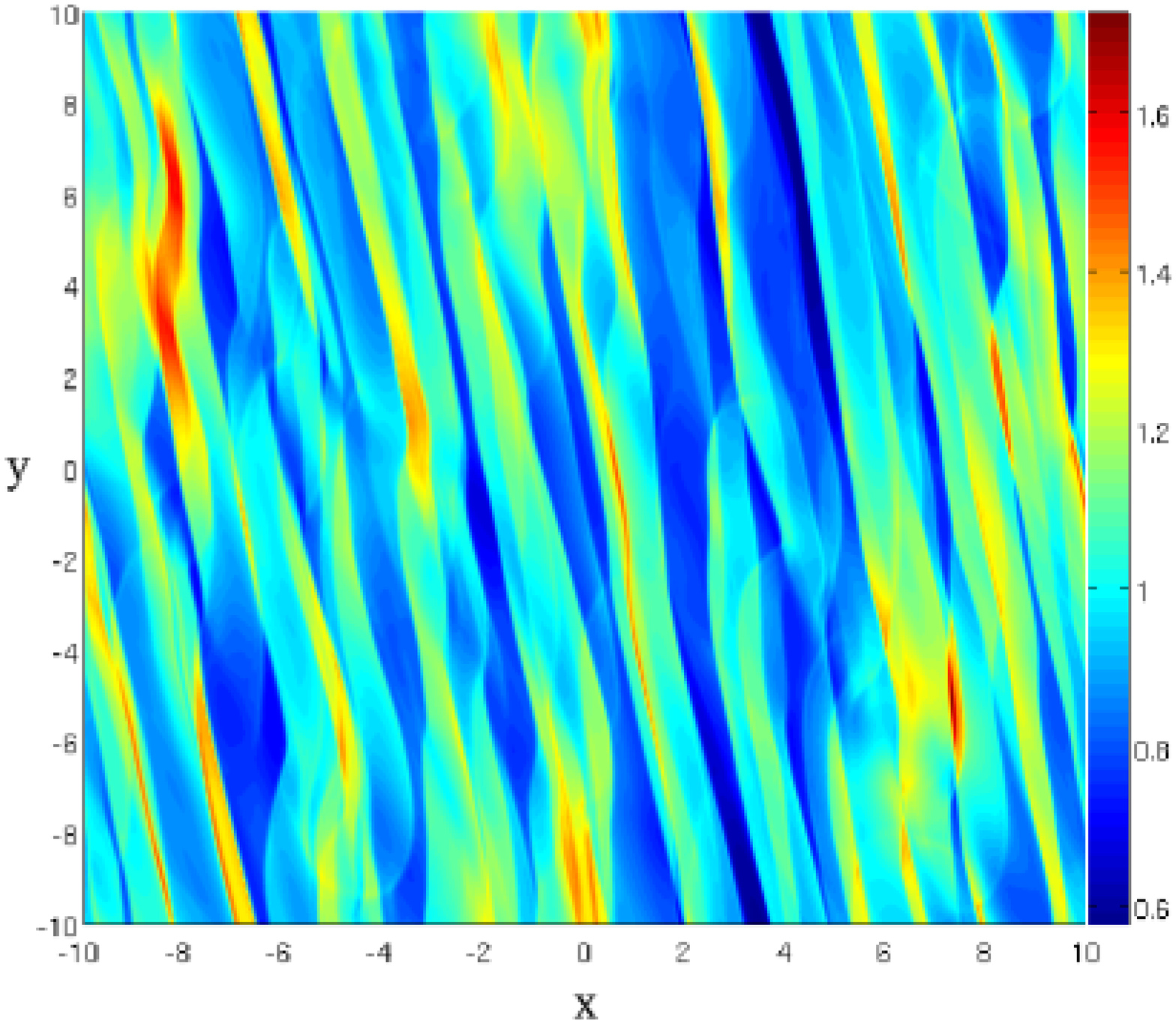}

\caption{Typical PV field (upper plot) at $t=44$ for the isothermal
fiducial model with self-gravity switched off. The sound speed is
equal to the average sound speed in the gravitoturbulent state (see
Fig.~3). Also shown is the corresponding surface density field
(lower plot). It is hard to see traces of individual vortices in the
surface density field; only shocks shed by these vortices are
visible. Vortices are organised into well-shaped larger scale
structures as opposed to what is observed in the self-gravitating
case in Fig.~4.}
\end{figure}

Since $Q\leq 1$ at the location of the stronger overdense regions,
self-gravity becomes dominant in the dynamics of vortices and
appears to strongly shear and deform the latter over about
$1-3\Omega^{-1}$ and most importantly to inhibit their further
growth in size. In other words, self-gravity opposes an inverse
cascade of energy to smaller wavenumbers and scatters it to higher
wavenumbers. This fact is also reflected in the autocorrelation
function of PV shown below (Fig.~6). The vortices can only grow to a
size comparable to the local Jeans scale $\lambda_J\sim Qc_s$ (in
non-dimensional variables). When the vortex size approaches the
local Jeans scale, it has no clear centre with high (by absolute
value) negative PV region surrounded by smaller (by absolute value)
PV region and, therefore, corresponds only to the stronger overdense
region (yellow and red regions in the surface density field in
Fig.~4) with no underdense centre. At this time the shape of the
vortex is more irregular, sheared and diffuse characterized by about
10 times smaller (by absolute value) PV than that of growing
vortices. From the pressure/internal energy field it is clear that
in these stronger overdense regions the internal energy is higher as
well and it is expected that corresponding $Q$ will eventually rise
and switch off self-gravity/gravitational instability at that
location. Then the overdense region will quickly disperse, or get
sheared away. After that sheared PV regions with higher $Q$, in
turn, can start another cycle of vortex formation by undergoing the
same nonlinear Kelvin-Helmholtz instability and again wrap up into
vortices. Thus, in self-gravitating discs vortices have recurring
nature -- each vortex forms, grows to a size of the order of the
local Jeans scale and after that gets sheared and destroyed by the
combined effects of self-gravity and Keplerian shear. Each such a
cycle typically lasts for about 2 orbital periods or less (1 orbital
period$=2\pi/\Omega$). Then sheared PV regions turn into vortices
and after that the whole process starts again. The cooling ensures
that the minimum $Q_{min}$ is kept low ($0.6-0.7$) and nearly
constant with time, so that self-gravity continues to play a role.

Above we have described the behaviour of some individual vortices
only. From Fig.~4 we see that the PV field contains vortices with
different sizes that never get organised into distinct coherent
vortices as they do in the non-self-gravitating case (Fig.~5).
Vortices evolve differently -- some of them are at the end of
evolution having already grown to the Jeans scale sizes and are
characterized by clearly identifiable overdense regions and low $Q$,
but others have yet to go through this phase and, therefore, may
still have underdense central regions surrounded by higher density
regions and correspond to higher $Q$. Thus, regions of high and low
$Q$ coexist throughout the disc at all times. A similar
gravitoturbulent state was observed in the global disc simulations
by \citet{WMN02} in the context of ISM turbulence. However, these
authors did not measure PV corresponding to high and low $Q$
regions. In general, the overall dynamical picture, as mentioned, is
very irregular and it is hard to keep track of individual vortices
for several orbital periods, because they are short-lived
structures. Figure~5 shows the stark contrast between
non-self-gravitating and self-gravitating cases. In this figure we
present a snapshot of the same fiducial model, but without
self-gravity and with constant sound speed (isothermal equation of
state) equal to the average sound speed of the fiducial model in the
gravitoturbulent state (see Fig.~3). In this case the dynamical
picture is identical to that described by other authors
\citep[e.g.,][]{UR04,JG05,SSG06}. Now it is easier to trace each
vortex and to see how they merge into larger vortices. In the
self-gravitating case, irregular and chaotic phases of vortex
evolution is quasi-steady, or equivalently self-sustained during the
entire course of evolution, whereas in the non-self-gravitating
case, though vortices are well-organised and regular, they gradually
decay.

\begin{figure}
\includegraphics[width=\columnwidth]{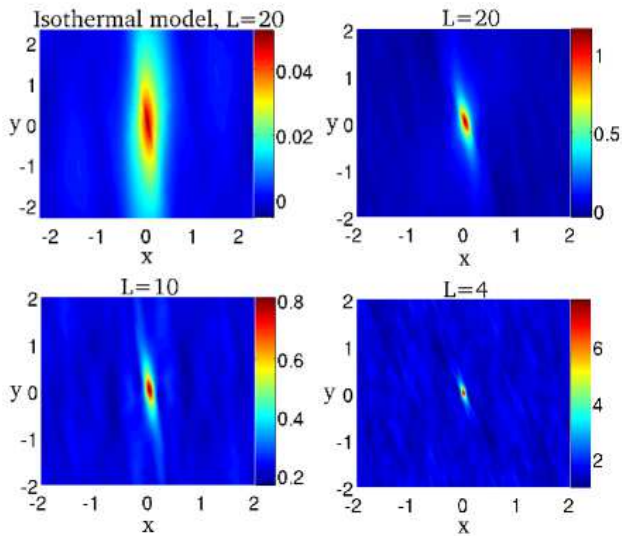}
\caption{Autocorrelation functions of PV for four models: isothermal
non-self-gravitating with $L=20$ (Fig.~5) and three self-gravitating
with $L=20, 10, 4$ in the gravitoturbulent state. For the two
fiducial models with the same $L=20$ the correlation length is
largest for the isothermal non-self-gravitating model and greatly
decreases in the presence of self-gravity. It also decreases with
decreasing $L$.}
\end{figure}

We also carried out simulations for two other models with sizes
$L=4$ and $L=10$ and with the same type of initial conditions. The
dynamical picture of vortex evolution described in detail for the
fiducial model remains qualitatively the same. There are differences
only in saturation times, and (average) values of various quantities
in the gravitoturbulent state. In Fig.~6 we compare the
autocorrelation functions of PV for these models with that of the
non-self-gravitating isothermal model plotted in Fig.~5. The extent
of the autocorrelation function, or the correlation length, remains
on average unchanged with time in the gravitoturbulent state and
does not depend on the spectrum of initial conditions. For the given
model size $L=20$, the correlation length is smaller than that in
the non-self-gravitating case implying that vortices are less
coherent in the gravitoturbulent state (we remind the reader that
the sound speed of the isothermal model, determining the correlation
length for this model, is equal to the average sound speed of the
fiducial self-gravitating model). This again can be explained by the
tendency of self-gravity to oppose inverse cascade of energy to
larger scales by scattering it to smaller scales and thereby
broadening the spectrum of the PV autocorrelation function. An
alternative explanation is that in the self-gravitating case the
contribution of Jeans scale size PV regions in the correlation
function is smaller than that of smaller size PV regions. The
correlation length appears to decrease with decreasing model size,
whereas in the isothermal case it does not depend on the size of the
model and is determined by the disc scale height \citep{JG05}. In
any case, correlation length is always smaller than the model size
justifying the local treatment of vortices.

\section{Summary and Discussions}

We have studied the specific properties of potential vorticity (PV)
evolution in self-gravitating discs in the shearing sheet
approximation. The evolution has been traced via numerical
integration of the basic hydrodynamic equations supplemented with
Poisson's equation taking care of self-gravity. Since we are
interested particularly in the properties of vortex evolution in
quasi-steady gravitoturbulence, we have chosen a simple cooling law
with a constant cooling time large enough so that the disc settles
down into this state. Our analysis has shown that in the
self-gravitating case, vortices appear as transient structures
undergoing recurring phases of formation, growth to sizes comparable
to a local Jeans scale, and eventual shearing and destruction due to
the combined action of self-gravity (gravitational instability) and
background Keplerian shear. Each such a phase lasts for a few
orbital periods. As a result, the overall dynamical picture is
irregular, consisting of many transient vortices at different
evolutionary stages and, therefore, with various sizes up to the
local Jeans scale. By contrast, in the non-self-gravitating case,
long-lived vortices form and grow in size via merging into larger
ones until eventually their size reaches the disc scale height. The
motion within vortices, or more precisely in the vicinity of
negative PV regions, is a complex mixture of compressive (wave) and
vortical motions which are difficult to separate from each other.
Compressive motions turn into shocks afterwards, or equivalently,
vortices appear to generate shocks. Therefore, the dynamics of
compressive motions (density waves) and vortices are coupled
implying that, in general, one should consider both vortex and
spiral density wave modes to get a proper understanding of
self-gravitating disc dynamics.

It is well known that overpressure/overdensity regions, if present
in a disc, can act as traps for dust particles
\citep[e.g.,][]{HB03,Rietal04,Rietal06,FN05,Lyraetal08}. As has been
demonstrated in our simulations, in self-gravitating discs
anticyclonic vortices, or generally negative PV regions, are able to
produce quite noticeable overpressure/overdensities, though not
long-lived. They have a transient character and vary on the
dynamical timescale. Hence, given such an irregular and rapidly
changing nature of vortex evolution in self-gravitating discs, it
seems difficult for corresponding overdensities/overpressures to
effectively trap dust particles in their centres. Further study of a
coupled system -- dust particles embedded in a self-gravitating
gaseous disc -- is, however, required to strengthen this conclusion.

Here we have considered the simplified case of a razor-thin (2D)
disc in order to gain first insight into the effects of self-gravity
on vortex dynamics. Obviously, for a fuller understanding 3D
treatment is necessary. The situation can be different in the 3D
self-gravitating case, which is evidently stratified in the vertical
direction. \citet{BM05} demonstrated that in the 3D
non-self-gravitating stratified shearing box, off-midplane vortices
appear and survive for many orbital periods. A similar 3D study by
\citet{SSG06}, but not including stratification, showed that
vortices are unstable and get quickly destroyed due to an elliptical
instability. So, it can be said that the presence of stratification
helps the survival of vortices. On the other hand, as we have seen
here, self-gravity does not favour long-lived vortices. But in the
3D case the effect of self-gravity is somewhat reduced compared with
that in the 2D case considered here. Thus, in a 3D generalization of
the present problem there are two competing factors --
stratification and self-gravity -- and it remains to be seen in
numerical experiments which of these two prevails. Another question
of interest in the 3D case (either self-gravitating or not) is, as
pointed out also by \citet{JG05}, if long-lived off-midplane
vortices can be generated from a random PV distribution, as happens
in the 2D case, rather than from specially chosen vortex solutions
as in simulations of \citet{BM05}.

\section*{Acknowledgments}
G.R.M. would like to acknowledge the financial support from the
Scottish Universities Physics Alliance (SUPA). He also thanks G. D.
Chagelishvili and A. G. Tevzadze for helpful discussions and for
reading the manuscript. The original version of the ZEUS code suited
for the shearing sheet as well as useful comments and suggestions
regarding numerical simulations of vortices in discs were kindly
provided by C. F. Gammie.

\end{document}